\documentclass[12pt]{iopart}

\usepackage{cite}
\usepackage{color,soul}
\usepackage{iopams}
\expandafter\let\csname equation*\endcsname\relax
\expandafter\let\csname endequation*\endcsname\relax
\usepackage{amsmath}
\usepackage{tensor}
\usepackage{appendix}
\usepackage[colorlinks=true,linkcolor=blue,citecolor=blue,urlcolor=blue]{hyperref}
\newcommand{\eqr}{\eqref}

\newcommand{\R}{\mathbb{R}}
\newcommand{\C}{\mathbb{C}}

\newcommand{\E}{\mathcal{E}}
\newcommand{\mueff}{\mu_{\text{eff}}}
\newcommand{\aint}{a_{\text{int}}}
\newcommand{\beq}{\begin{equation}}
\newcommand{\eeq}{\end{equation}}

\usepackage{tabularx} 
\usepackage{amsmath,amssymb,amsfonts,tensor,physics,overarrows,empheq}  
\usepackage[normalem]{ulem} 
\usepackage{amssymb,latexsym}
\usepackage{amsmath}
\usepackage{tensor}
\usepackage[utf8]{inputenc}
\usepackage[colorlinks=true,linkcolor=blue,citecolor=blue,urlcolor=blue]{hyperref}
\usepackage{graphicx}     
\usepackage[dvipsnames]{xcolor}
\usepackage{floatflt}     
\usepackage{wrapfig}
\usepackage{color}
\usepackage{pstricks}
\usepackage{etoolbox}
\usepackage{subcaption}
\usepackage{color}
\definecolor{DarkRed}{rgb}{0.35,0.01,0.01}
 \definecolor{Linen}{rgb}{0.98,0.98,0.94}
 \definecolor{Blue}{rgb}{0.,0.,1.0}
 \definecolor{DarkBlue}{rgb}{0.099,0.099,0.44}
 \definecolor{DarkGreen}{rgb}{0.0,0.4,0.0}
 \definecolor{Turquoise}{rgb}{0.0,0.9,0.7}
 \usepackage{tikz}
\usepackage{tkz-euclide}
\usepackage{tikz-3dplot}
\usepackage{pgfplots,pgfplotstable}
\usepackage[makeroom]{cancel}
\pgfplotsset{compat=newest}
\usetikzlibrary{shapes,shapes.arrows,shapes.symbols,shapes.geometric, calc,positioning}
\usepackage{mathtools}
\tikzstyle{start} = [rectangle, rounded corners, 
minimum width=3cm, 
minimum height=0.7cm,
text centered, 
text width=3cm, 
draw=black, 
fill=orange!30]

\tikzstyle{goal2} = [rectangle, rounded corners, 
minimum width=3cm, 
minimum height=0.7cm,
text centered, 
text width=3cm, 
draw=black, 
fill=red!30]

\tikzstyle{goal1} = [rectangle, rounded corners, 
minimum width=3cm, 
minimum height=0.7cm,
text centered, 
text width=3cm, 
draw=black, 
fill=purple!30]

\tikzstyle{goal3} = [rectangle, rounded corners, 
minimum width=3cm, 
minimum height=0.7cm,
text centered, 
text width=2.5cm, 
draw=black, 
fill=green!30]
\tikzstyle{arrow} = [thick,->,>=stealth]

\makeatletter
\newcommand{\mainmatter}{%
  \setcounter{footnote}{0}%
  \patchcmd{\@makefntext}{\fnsymbol}{\arabic}{}{}%
  \patchcmd{\@thefnmark}{\fnsymbol}{\arabic}{}{}%
  \def\@makefnmark{\textsuperscript{\arabic{footnote}}}%
}
\makeatother

\begin{document}

\title[Superradiance in CWG]{Superradiant Suppression of Non-minimally Coupled Scalar fields for a Rotating Charged dS Black Hole in Conformal Weyl Gravity}

\author{Owen Gartlan$^{1}$, Jacob March$^{2}$, Leo Rodriguez$^{1,3}$, Shanshan Rodriguez$^{1,3,4,*}$, Yihan Shen$^{1}$}

\address{$^1$ Department of Physics, Grinnell College, Grinnell, IA, 50112, USA}
\address{$^2$ Department of Physics, Northeastern University,  Boston, MA, 02115, USA}
\address{$^3$ Department of Physics, Worcester Polytechnic Institute, Worcester, MA, 01609, USA}
\address{$^4$ Center for Astrophysics, Harvard \& Smithsonian, Cambridge, MA 02138}
\address{$^*$ Author to whom any correspondence should be addressed.}
\ead{rodriguezs@grinnell.edu}
\vspace{10pt}
\begin{indented}
\item[]\today
\end{indented}

\begin{abstract}

In this study, we present an analytical investigation of the superradiant scattering of a massive charged conformally coupled scalar field in rotating charged $de~Sitter$ black hole spacetimes within two gravitational theories: General Relativity (GR) and fourth–order Conformal (Weyl–squared) Gravity (CWG). For the massless charged conformally coupled scalar, we exploit a recently discovered correspondence between the Heun equation and the semiclassical limit of Belavin-Polyakov-Zamolodchikov (BPZ) equations in two-dimensional conformal field theory to solve for the superradiant amplification factors as controlled expansions in a small parameter scaling. For the massive charged conformally coupled scalar, we use WKB methods to derive an order of magnitude approximation for the amplification factors in the cosmological region in terms of those in the region $r_+\ll r \ll r_c$ where $r_+$ and $r_c$ are the outer and cosmological event horizons, respectively. For both the massless and massive sectors, suppression of superradiant amplification in CWG relative to that in GR is observed across the parameter regimes studied. Particularly, in the massive sector, we find strong exponential suppression of superradiant amplification on the order of $e^{-2\mu\Lambda^{-1/2}}$ in the cosmological region.

\end{abstract}
\newpage 
%
%
%
%
\mainmatter
\section{Introduction}
The astrophysical reality of black holes has now been confirmed through multiple observational channels, most notably by the detection of gravitational waves from compact-binary mergers and the horizon-scale imaging of the supermassive black holes in M87 and Sgr A* \cite{LIGOScientific:2016aoc, LIGOScientific:2019fpa, EventHorizonTelescope:2019ths, EventHorizonTelescope:2019dse, EventHorizonTelescope:2022wkp, EventHorizonTelescope:2022wok}. These breakthroughs have transformed black holes from purely theoretical solutions of general relativity into directly observed astrophysical objects, which may serve as powerful cosmic engines of extractable energy \cite{Penrose:1969pc, Misner:1972kx}. Understanding how such energy can be tapped is therefore of both fundamental and astrophysical importance. 

Superradiant scattering, the amplification of bosonic waves through the extraction of energy and angular momentum from a rotating or charged black hole, is a fundamental phenomenon in black-hole perturbation theory and an important probe of horizon stability and energy-extraction mechanisms \cite{Brito:2015oca}. Following the original Penrose process \cite{penroseExtractionRotationalEnergy1971, Penrose:1969pc, Penrose:1971uk} and early analysis of wave amplification and resonant scattering \cite{damourQuantumResonancesStationary1976}, superradiance has been extensively studied in Kerr, Kerr-Newman, and Reissner-Nordstr\"om black holes within General Relativity (GR) \cite{DiMenza:2014vpa, Benone:2014qaa, Benone:2015bst, Balakumar:2020gli, Glampedakis:2001cx, Dolan:2008kf, Benone:2019all, Mascher:2022pku, wangExtractingEnergyMagnetic2022, yuanConstraintsUltralightScalar2022}. These studies span classical energy-extraction mechanisms, superradiant instabilities, and observational constraints on light bosonic fields derived from black-hole spin measurements and gravitational-wave observations. In GR, a typical superradiant condition for a bosonic mode of frequency $\omega$, azimuthal number $m$, and field charge $q$, is $\omega<m\Omega_H+q\phi_H$, where $\Omega_H$ and $\phi_H$ are the angular velocity and electric potential of the event horizon. In this regime, the reflected flux exceeds the incident flux, yielding a positive amplification factor $Z_{lm}$ and signaling net energy extraction from the black hole. If a superradiantly amplified mode is reflected back toward the black hole by an effective ``mirror", it can be repeatedly amplified. The resulting exponential growth of the trapped mode leads to a superradiant instability commonly referred to as the black hole bomb \cite{Berti:2019wnn, Press:1972zz, Cardoso:2004nk,Cardoso:2013krh, Herdeiro:2013pia, Dolan:2015dha,  Dias:2018zjg}. 

The superradiant condition can depend sensitively on the asymptotic structure of the spacetime as well as on the underlying gravitational theory. In particular, black holes in asymptotically flat, $de~Sitter$ ($dS$), or $Anti$-$de~ Sitter$ ($AdS$) backgrounds may exhibit qualitatively different superradiant behaviour because the horizon properties, boundary conditions, and effective trapping mechanisms are modified by the global geometry (the reader can consult the review \cite{britoSuperradiance2020Edition2020} for a broad study of the vast literature in superradiance). Likewise, for black holes arising in theories beyond general relativity, the superradiant threshold can be altered by changes in the metric, additional charges or fields, nonminimal couplings, or modified horizon dynamics. As a result, superradiance provides a valuable probe of strong-gravity physics: its amplification conditions, instability spectra, and observational signatures may help distinguish general relativity from alternative theories and potentially reveal imprints of new physics, including effects that could encode aspects of quantum gravity.

In asymptotically $dS$ spacetimes, the presence of a cosmological horizon provides a natural outer boundary for defining conserved fluxes and scattering amplitudes. Rotating and charged $dS$ black holes exhibit particularly rich superradiant behavior for charged scalar fields, governed by the combined effects of rotation, electromagnetic coupling, and the multi-horizon structure of the geometry \cite{suzukiPerturbationsKerrdeSitter1998, mascherChargedBlackHoles2022}. At the same time, the existence of multiple horizons complicates analytic treatments, as the associated radial equations typically possess several regular singular points and nontrivial global boundary conditions.

Analytical approaches to superradiance, therefore, depend sensitively on the spacetime geometry and field content. In asymptotically flat or $AdS$ backgrounds, low-frequency amplification and instability growth rates are often obtained using matched asymptotic expansions between near-horizon and far-region solutions \cite{Green:2015kur, Li:2019tns, Yang:2022uze, Chen:2021zqs, Cuadros-Melgar:2021sjy, Khodadi:2021mct, Richarte:2021fbi, Ishii:2022lwc}. In spacetimes with multiple horizons, such as $dS$ geometries, the separated radial equations frequently reduce to Fuchsian differential equations with four or more regular singular points, including Heun-type equations, whose connection problems encode the scattering amplitudes \cite{Zhu:2014sya, Destounis:2019hca, Mascher:2022pku}. Exact solutions are generally available only in restricted regimes, motivating perturbative treatments in small frequency or small horizon-separation limits. 

In addition, superradiant scattering and quasinormal-mode spectroscopy are closely related aspects of the same underlying perturbation problem \cite{Zouros:1979iw}. In both cases, one studies the spectrum of the separated radial equation subject to physically motivated horizon and asymptotic boundary conditions; superradiant instabilities arise when trapped or quasibound modes satisfy the superradiant threshold and develop a positive imaginary part of the frequency \cite{Berti:2009kk}. Consequently, several of the standard techniques used in QNM studies, such as matched asymptotic expansions, continued-fraction methods, and WKB analysis, also play a central role in superradiance \cite{Konoplya:2011qq}. In particular, for massive fields or more intricate effective potentials, semiclassical WKB methods are commonly employed to analyze barrier penetration, quasibound states, resonance frequencies, and instability conditions, with amplification or decay governed by tunneling across classically forbidden regions \cite{Zouros:1979iw, Xia:2023zlf}. These techniques provide complementary analytic control over superradiant scattering while highlighting its sensitivity to horizon structure and global geometry \cite{Iyer:1986np}.

While most existing work has focused on black holes within GR, comparatively little is known about superradiant scattering in rotating and charged black holes arising in alternative theories of gravity. Conformal Weyl gravity (CWG) and fourth order gravity theories in general \cite{Brensinger:2017gtb,Brensinger:2020gcv} are of great recent interest as helpful tools in gravitational/cosmological physics \cite{Brensinger:2019mnx,Mannheim:2009qi,Edery:1997hu,Edery:1998zi,Edery:2001at}, black hole physics \cite{Zou:2020rlv,Momennia:2019cfd,Momennia:2019edt,Momennia:2018hsm}, their thermodynamics \cite{Xu:2018liy,Bambi:2017ott} and role in quantum gravity \cite{Biedke:2026job,RRW,Mannheim:2009qi,Stelle:1976gc,Edery:2006hg}. Vacuum CWG is given by the Weyl tensor squared action:
\begin{align}\label{eq:CWGA}
S_{CWG}=\alpha_{c}\int d^4x\sqrt{-g}C^{\alpha\mu\beta\nu}C_{\alpha\mu\beta\nu},
\end{align}
which is (similarly to $SU(N)$ gauge theories in four dimensions) a diffeomorphism and conformally invariant theory \cite{Mannheim:2011ds,Kiefer:2017nmo} for unit-less coupling $\alpha_{c}$. In addition, CWG includes all of the Einstein-Hilbert (EH) vacuum black hole solutions as part of its solution space and cosmological dynamics appear naturally within this formalism \cite{Mannheim:1988dj,Mannheim:2005bfa}. Additionally, several Einstein-Hilbert non-vacuum black hole solutions have analogue counterparts within CWG \cite{Mannheim:1990ya}. The above leads to specific regimes where the two theories may be considered equivalent \cite{Maldacena:2011mk,Anastasiou:2016jix,Anastasiou:2020mik}. Thus, CWG provides an important and interesting alternative (to GR) playground, which gives exact rotating and charged black-hole solutions whose radial structure differs from the Kerr-Newman family through a nontrivial dependence on electric charge \cite{mannheimSolutionsReissnerNordstromKerr1991, liuChargedRotatingAdS2013}. These modifications alter horizon locations, ergoregions, and the effective potentials experienced by perturbing fields. Since superradiant amplification depends sensitively on these geometric features, it provides a natural diagnostic for assessing how departures from GR affect black-hole scattering and stability properties.

In this work, we study superradiant scattering of a charged scalar field in two classes of rotating, charged $de~Sitter$ black holes: the Kerr-Newman-$de~Sitter$ ($KNdS$) solutions of GR and a Kerr-Newman-like solution of conformal Weyl gravity \cite{liuChargedRotatingAdS2013}, which we denote $KNdSCG$. For a massless conformally coupled charged scalar field, the separated radial equation possesses four finite regular singular points associated with the black-hole and cosmological horizons, together with a removable regular singularity at infinity. The equation can therefore be reduced to the general Heun form. Although Heun equations arise frequently in black-hole perturbation theory \cite{suzukiPerturbationsKerrdeSitter1998, baticHeunEquationTeukolsky2007, hortacsuHeunFunctionsTheir2018}, their connection problems are generally intractable.

Recent developments have shown that, in the small crossing-ratio regime, the connection problem for Heun equations can be treated perturbatively by exploiting its relation to the semiclassical limit of Belavin-Polyakov-Zamolodchikov (BPZ) equations in two-dimensional conformal field theory. Within this framework, the relevant connection coefficients are expressed in terms of semiclassical conformal blocks and hypergeometric connection matrices \cite{lisovyyPerturbativeConnectionFormulas2022, bonelliIrregularLiouvilleCorrelators2023, aminovBlackHolePerturbation2023}. In the massless, conformally coupled case, we adopt this approach to compute reflection and transmission amplitudes for charged scalar perturbations, from which the superradiant amplification factor $Z_{lm}$ is obtained in a controlled small crossing-ratio expansion. When the scalar field is massive, the regular singularity at infinity becomes non-removable, and the radial equation no longer admits a reduction to Heun form. In this regime, we instead analyze the problem using semiclassical WKB methods. By examining the associated effective potential, we identify a near-horizon propagation region separated from the cosmological horizon by an evanescent barrier. The transmission of flux across this region is characterized by a WKB tunneling action $S$, yielding an exponential suppression factor of the form $e^{-2S}$. This approach provides analytic control over superradiant scattering in parameter regimes where exact or perturbative special-function techniques are no longer applicable.
 
The paper is organized as follows. In Section \ref{sect:Geodesics}, we introduce the Kerr-Newman $de~Sitter$-like black hole in Conformal Weyl Gravity and compare its  horizon and ergoregion structure with that of the Kerr-Newman $de~Sitter$ spacetime in General Relativity. In Section \ref{sect:NMCoupled}, we motivate the non-minimally coupled Klein-Gordon Equation and the choice of $\xi=1/6$ for the dimensionless coupling constant in four spacetime dimensions. In Section \ref{sect:ACCSSP}, we derive the separated equations for a charged, massive, conformally coupled scalar field and establish the associated boundary flux relations. Section \ref{sect:SoaMAF} develops the perturbative Heun-CFT approach for a massless charged conformally coupled scalar field and applies it to compute analytic expressions for the superradiant amplification factors. In Section \ref{sect:SoaMiveSF}, we present a semiclassical WKB analysis for a massive charged conformally coupled scalar field and estimate the corresponding transmission suppression across the cosmological barrier. Finally, Section \ref{DanC} summarizes our results, discusses the limitations of the analytic approximations employed, and outlines directions for future numerical and nonlinear investigations. We have assumed geometrized units $\hslash=c=G=4\pi \varepsilon_0=1$.

\section{Geodesics of a Charged Rotating de Sitter Black Hole in Conformal Weyl gravity} \label{sect:Geodesics}
The charged rotating $de~Sitter$ black hole we consider is a solution in Conformal Weyl Gravity (CWG), with an action minimally coupled to the Maxwell Field given by
\beq\label{action}
S=\alpha \int d^4x\sqrt{-g}\left( \frac{1}{2} C^{\mu \nu \rho \sigma} C_{\mu \nu \rho \sigma} + \frac{1}{3} F^2 \right),
\eeq
where $\alpha$ is the CWG coupling constant, $F=dA$ is the strength of the Maxwell field, and $C_{\mu \nu \rho \sigma}$ is the canonical Weyl tensor \cite{liuChargedRotatingAdS2013}. The equations of motion derived from the above action \eqref{action} are
\begin{equation}\label{EOM}
\nabla^{\mu}F_{\mu\nu}=0, \quad -\alpha(2\nabla^{\rho}\nabla^{\sigma}+R^{\rho\sigma})C_{\mu\rho\sigma\nu}+\frac{2}{3}\alpha(F_{\mu\nu}-\frac{1}{4}F^2g_{\mu\nu})=0,
\end{equation}
which results in the following exact, asymptotically $de~Sitter$ black hole solution \cite{mannheimSolutionsReissnerNordstromKerr1991,liuChargedRotatingAdS2013} \footnote{The form of the metric in \eqref{metric} differs from that found in Ref. \cite{liuChargedRotatingAdS2013} by the temporal diffeomorphism $t\rightarrow \frac{t}{\Xi}$.}. 
\begin{equation}\label{metric}
\begin{split}
ds^2=\rho^2\left(\frac{dr^2}{\Delta_r}+\frac{d\theta^2}{\Delta_{\theta}}\right)+& \frac{\Delta_{\theta}\sin^2\theta}{\Xi^2\rho^2}\left(adt-(r^2+a^2)d\phi\right)^2-\frac{\Delta_r}{\Xi^2\rho^2}\left(dt-a\sin^2\theta \ d\phi\right)^2, \\
& dA= -\frac{Qr}{\Xi \rho^2}\left(dt-a\sin^2\theta \ d\phi\right),
\end{split}
\end{equation}
where
\begin{equation}\label{metric functions}
\begin{split}
& \rho^2=r^2+a^2\cos^2\theta,\quad \Delta_{\theta}=1+\frac{1}{3}\Lambda a^2 \cos^2\theta,\quad \Xi=1+\frac{1}{3}\Lambda a^2,\\
& \Delta_r=\left(r^2+a^2\right)\left(1-\frac{1}{3}\Lambda r^2\right)-2Mr+\frac{Q^2 r^3}{6M}.
\end{split}
\end{equation}
Here, $Q$, $a$ and $\lambda$ are the black hole charge, spin, and the cosmological constant, respectively. The Ricci Scalar for this metric is given by $R=4\Lambda -\frac{Q^2 r}{M\rho^2}$. For the remainder of this paper, we will refer to the above solution as the Kerr-Newman Conformal Weyl Gravity ($KNdSCG$) metric, which reduces to the regular Kerr $de~Sitter$ metric when $Q=0$. It differs from the regular Kerr-Newman dS ($KNdS$) metric only in the cubic dependence on $r$ in the $Q^2$ term of the metric function $\Delta_r$. 
\begin{figure}[ht]
\includegraphics[width=1\linewidth, height=6cm]{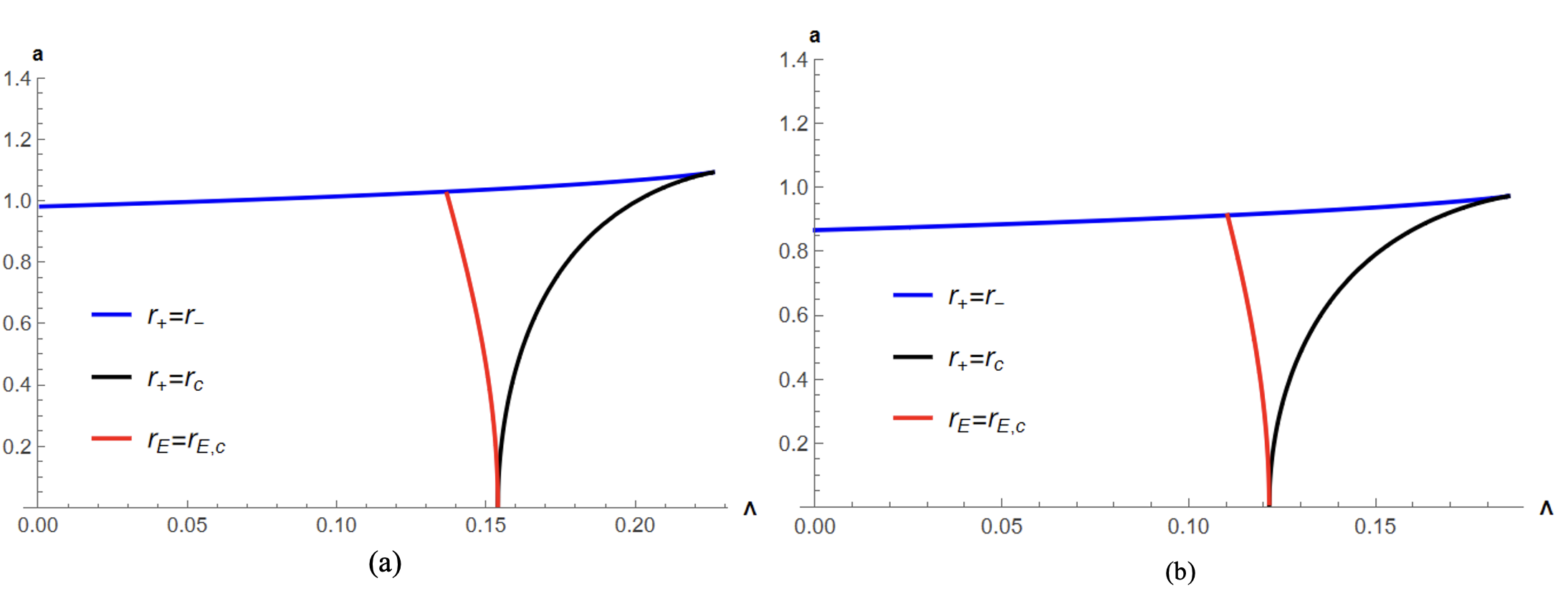} 
\caption{Parameter space (a, $\Lambda$) for the $KNdSCG$ (Panel (a)) and $KNdS$ (Panel (b)) spacetimes. The region bounded by the blue and red curves in the lower left corner denotes the black hole region with at least one event horizon. The red curve indicates $r_E=r_{E,c}$ where the cosmological ergosphere and the black hole ergosphere coincide. To its right side, the black hole no longer possesses an ergosphere. For both cases: $Q=0.5$.}
\label{ExtCond}
\end{figure}
The horizons of the black hole are determined by 
\begin{equation}
\Delta_r=\left(r^2+a^2\right)\left(1-\frac{1}{3}\Lambda r^2\right)-2Mr+\frac{Q^2 r^3}{6M}=0,
\end{equation}
which generally has four roots. The three positive roots that we denote as  $r_{-}$, $r_{+}$, and $r_{c}$ from smallest to largest are the inner, outer, and cosmological event horizons, while the negative root $r_n$ is ignored. We study the extremal conditions for the parameter space ${\Lambda M^2, a/M}$ by requiring $r_{-} \le r_{+} \le r_{c}$, and compare such conditions to those for the regular Kerr-Newman dS ($KNdS$) black hole. As shown in Fig.~\ref{ExtCond}, both $r_{+}=r_{-}$ (blue) and the $r_{+}=r_{c}$ (black) curves are similar in shape and form a closed region. The $KNdSCG$ spacetime allows a slightly larger parameter space for both black hole spin $a$ and the cosmological constant $\Lambda$ compared to the $KNdS$ case for the same given black hole charge. 

\begin{figure}[ht]
\includegraphics[width=1\linewidth, height=6cm]{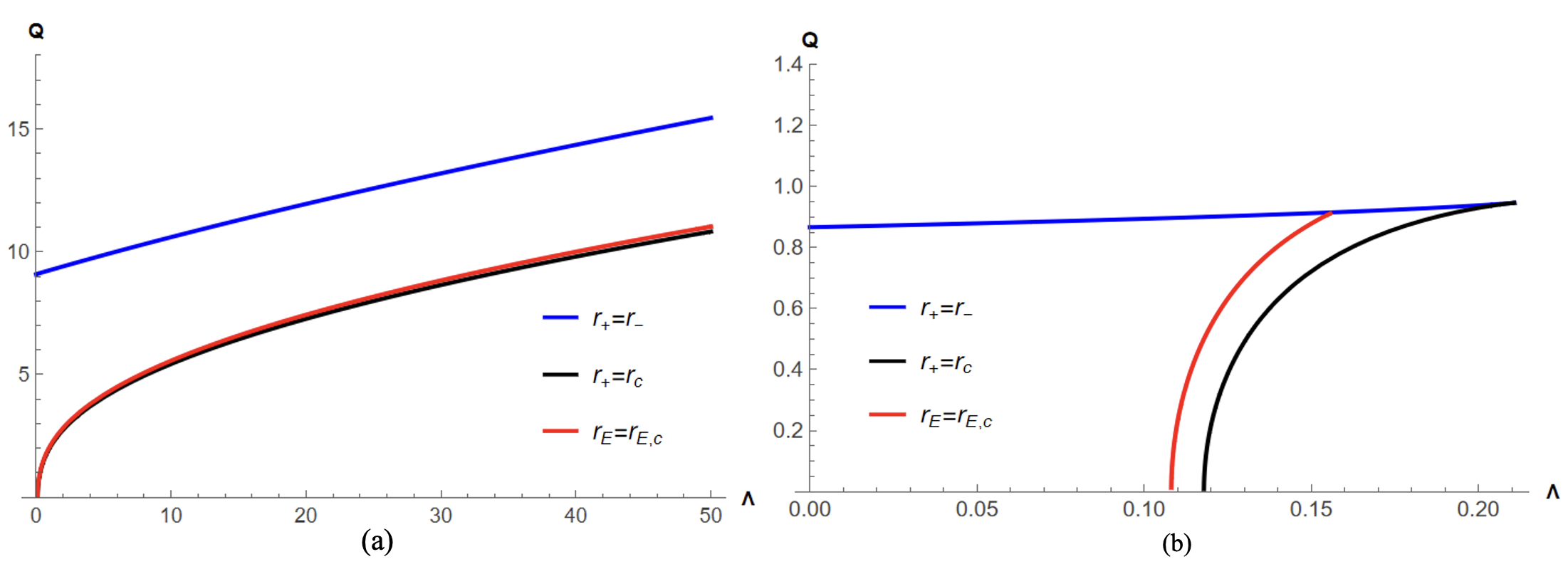} 
\caption{Parameter space (Q, $\Lambda$) for the $KNdSCG$ (Panel (a)) and $KNdS$ (Panel (b)) spacetimes. The region bounded by the blue and red curves in the lower left corner denotes the black hole region with at least one event horizon. The red curve indicates $r_E=r_{E,c}$ where the cosmological ergosphere and the black hole ergosphere coincide. To its right side, the black hole no longer possesses an ergosphere. For both cases: $a/M=0.5$.}
\label{ExtCond2}
\end{figure}

\begin{figure}[ht]
\includegraphics[width=1\linewidth, height=6cm]{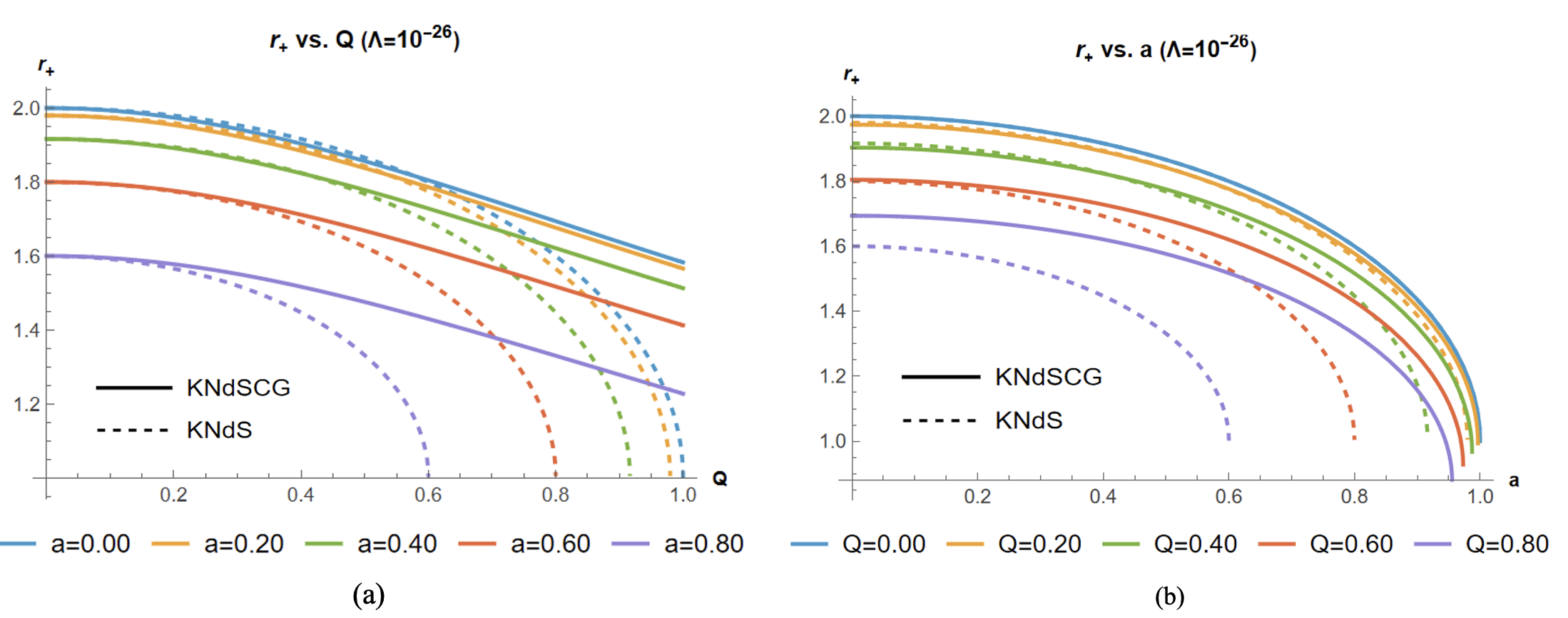} 
\caption{Outer event horizon, $r_{+}$, as a function of $Q/M$ (Panel (a)) and $a/M$ (Panel (b)). In both graphs, the \textbf{solid lines} represent the $KNdSCG$ spacetime and the \textbf{dashed lines} represent the $KNdS$ spacetime.}
\label{rplus}
\end{figure}

\begin{figure}[ht]
\includegraphics[width=1\linewidth, height=6cm]{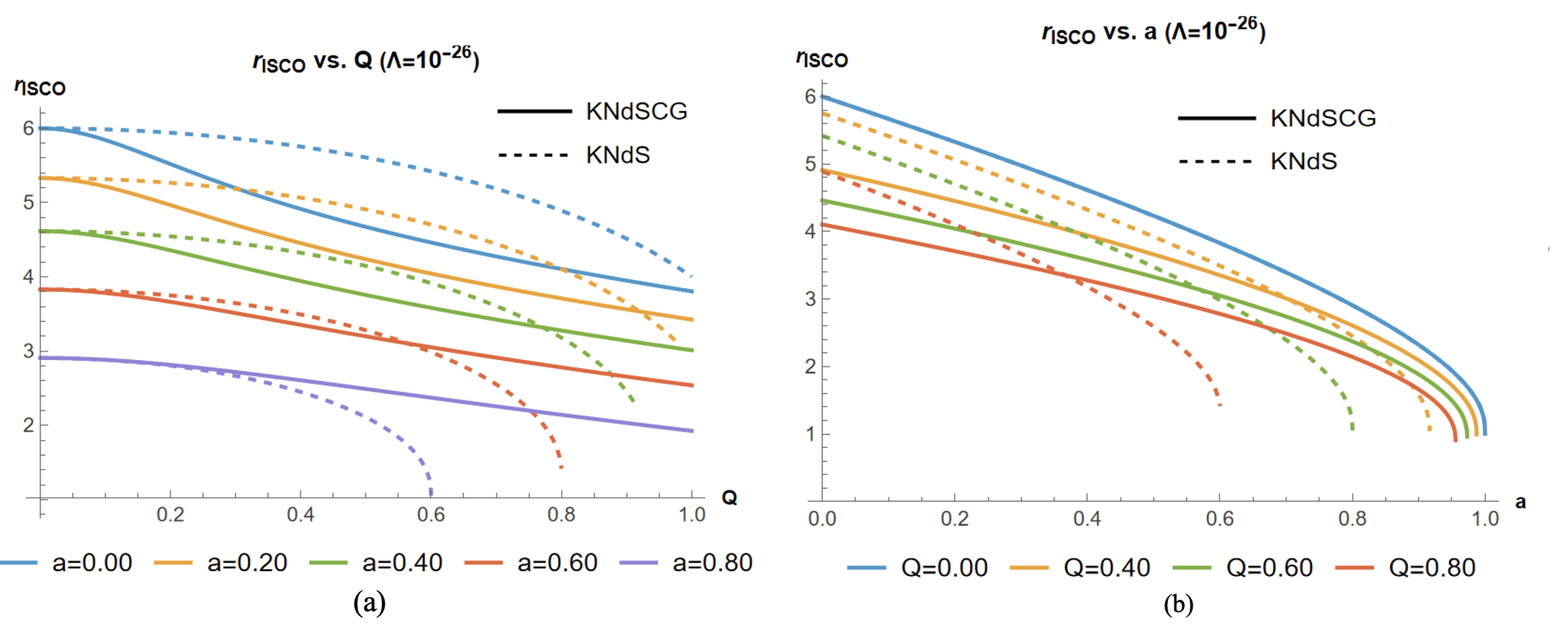} 
\caption{$r_{ISCO}$ as a function of $Q/M$ (Panel (a)) and $a/M$ (Panel (b)). In both graphs, the \textbf{solid lines} represent the $KNdSCG$ spacetime and the \textbf{dashed lines} represent the $KNdS$ spacetime.}
\label{rISCO}
\end{figure}

In contrast, Fig.~\ref{ExtCond2} shows a more substantial difference in the parameter space $(Q/M, \Lambda M^2)$ between these two spacetimes. For the $KNdSCG$ spacetime, the allowed region of $Q$ appears to be unbounded as $\Lambda$ varies for a fixed black hole spin, even within astrophysically realistic values of $\Lambda M^2 \sim 10^{-26}$. In comparison, the $KNdS$ spacetime exhibits a well-bounded range for both allowed black hole charge and cosmological constant.
Next, we compare these two spacetimes by examining their outer event horizon radius $r_{+}$ and innermost stable circular orbit (ISCO) across a range of parameters. For fixed spin $a$, both $r_{+}$ and $r_{ISCO}$ of the $KNdSCG$ solution closely track their $KNdS$ counterparts over a finite range of the black hole charge $Q$. As shown in Fig.~\ref{rISCO}(a), the ISCO radius begins to deviate from that of $KNdS$ at approximately $Q\sim 0.05M$. This range increases with spin, such that for $a/M=0.80$, the $KNdSCG$ and $KNdS$ ISCO radii remain comparable up to $Q \sim 0.2M$. In contrast, Fig.~\ref{rplus}(a) shows that the range of $Q$ over which the two spacetimes yield similar outer horizon radii decreases as $a$ increases.
Fig.~\ref{rplus} illustrates that $r_{+}$ increases monotonically with both $a$ and $Q$ for the $KNdSCG$ black hole, and that $r_{+}^{KNdSCG}$ generally exceeds $r_{+}^{KNdS}$ at larger values of $a$ and $Q$. Conversely, Fig.~\ref{rISCO} shows that the ISCO radius decreases monotonically with increasing $a$ and $Q$ for the $KNdSCG$ spacetime, with $r_{ISCO}^{KNdSCG}$ tending to be smaller than its $KNdS$ counterpart at larger $Q$ and smaller $a$. At sufficiently large values of $Q$, the functional dependence of both $r_{+}$ and $r_{ISCO}$ on $Q$ differs markedly between the two spacetimes, reflecting the distinct charge dependence of the $KNdSCG$ metric function $\Delta_{r}(r)$.
\begin{figure}[!htbp]
\includegraphics[width=1\linewidth, height=9cm]{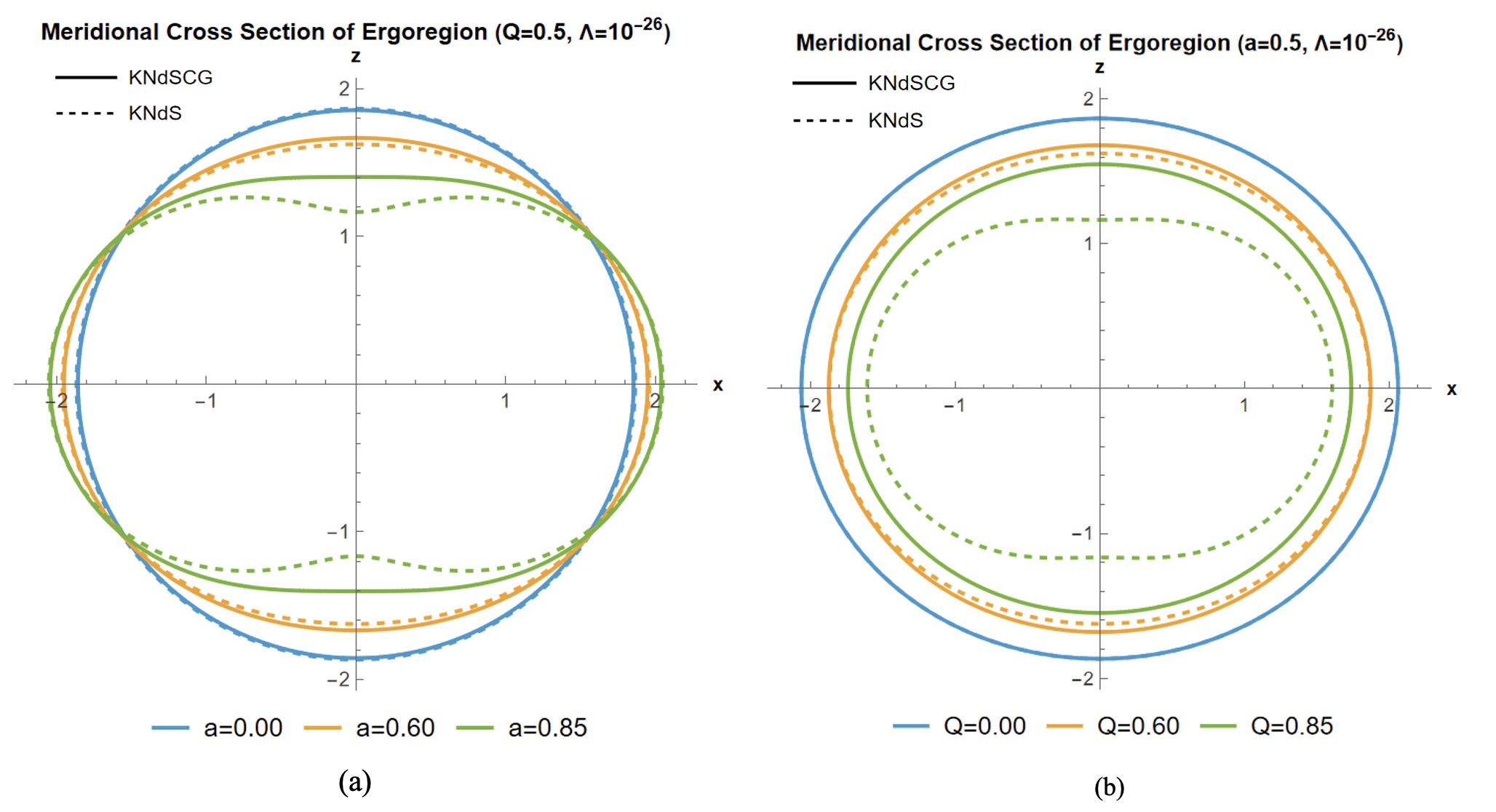} 
\caption{Meridional cross section of the ergoregion for varying $a/M$ (Panel (a)) and $Q/M$ (Panel (b)). In both graphs, the \textbf{solid lines} represent the $KNdSCG$ spacetime and the \textbf{dashed lines} represent the $KNdS$ spacetime.}
\label{Ergo}
\end{figure}

Finally, we consider the ergoregion, which plays a central role in black hole energy extraction mechanisms and is defined by the surface $g_{tt}=0$. As shown in Fig.~\ref{Ergo}, the ergoregion of the $KNdSCG$ spacetime is generally larger than that of $KNdS$. Its size increases with both $a$ and $Q$, as expected. However, the ergoregion of the $KNdSCG$ solution varies more gradually with these parameters than in the $KNdS$ case.

\section{Non-Minimally Coupled Klein-Gordon Equation}\label{sect:NMCoupled}
We consider a real scalar field $\phi$ of rest mass $\mu$ propagating in four spacetime dimensions. The most general, diffeomorphism-invariant quadratic action (up to total derivatives) for such a scalar contains a non-minimal coupling between the scalar field and the Ricci scalar \cite{faraoniConformallyCoupledInflation2013,santosQuasinormalModesMassive2021,calzaImportanceBeingNonminimally2025,stefanekLectureNotesNonminimally},
\begin{equation}\label{eq:action}
S[\phi] \;=\; -\frac{1}{2}\int_{\mathcal{M}} \! d^4x \,\sqrt{-g}\,\Big( g^{\mu\nu}\nabla_\mu\phi\,\nabla_\nu\phi + \mu^2\phi^2 + \xi R \phi^2 \Big)\,,
\end{equation}
where  $\xi\in\R$ is a dimensionless coupling constant. The choice $\xi=0$ corresponds to minimal coupling; in four spacetime dimensions, the value
\[
\xi_{\mathrm{conf}}=\frac{1}{6}
\]
is singled out as the conformal coupling for a classically massless scalar \cite{faraoniConformallyCoupledInflation2013,santosQuasinormalModesMassive2021,calzaImportanceBeingNonminimally2025,stefanekLectureNotesNonminimally}.

Variation of the action \eqref{eq:action} with respect to $\phi$ yields the generalized Klein--Gordon equation
\begin{equation}\label{eq:kg}
\big( \square - \mu^2 - \xi R \big)\phi = 0,
\qquad \square\equiv g^{\mu\nu}\nabla_\mu\nabla_\nu.
\end{equation}

\subsection{Green's function, Hadamard decomposition and short-distance behaviour}

Let $G_R(x',x)$ denote the retarded Green's function of the $\square$ operator appearing in \eqref{eq:kg}, defined by
\begin{equation}\label{eq:green}
\Big[ g^{\mu'\nu'}(x')\nabla_{\mu'}\nabla_{\nu'} - \mu^2 - \xi R(x') \Big] G_R(x',x) = -\delta(x',x),
\end{equation}
together with the retarded boundary conditions.  In a normal convex neighbourhood, the retarded Green's function admits the standard Hadamard decomposition \cite{dewittRadiationDampingGravitational1960}
\begin{equation}\label{eq:hadamard}
G_R(x',x) \;=\; \Sigma(x',x)\,\delta\!\big(\Gamma(x',x)\big) \;+\; W(x',x)\,\Theta\!\big(-\Gamma(x',x)\big),
\end{equation}
where $\Gamma(x',x)$ is Synge's world function (one-half the squared geodesic distance), $\delta(\Gamma)$ is the Dirac distribution supported on the past light cone of $x$, and $\Theta(-\Gamma)$ is the Heaviside distribution with support inside that cone.  The first term in Eq.~\eqref{eq:hadamard} propagates signals strictly on the light cone, while the second term encodes backscattering due to curvature and mass and represents propagation inside the light cone.

Requiring consistency with the Minkowski-space Green's function in the coincidence limit $x'\to x$ \cite{sonegoCouplingCurvatureScalar1993} yields the well-known short-distance expansions \cite{dewittRadiationDampingGravitational1960,sonegoCouplingCurvatureScalar1993}
\begin{align}
\Sigma(x',x) &= \frac{1}{4\pi} + O(x',x),\label{eq:sigma}\\[4pt]
W(x',x) &= -\frac{1}{8\pi}\Big[\mu^2 + \Big(\xi-\tfrac{1}{6}\Big)R(x)\Big] + O(x',x).\label{eq:W}
\end{align}
In flat spacetime $W_{\eta}(x',x)=-\mu^2/(8\pi)+O(x',x)$, and the above equation \eqref{eq:W} shows explicitly how curvature modifies the flat-space tail through the combination $\mu^2+(\xi-1/6)R$.

\subsection{Causality argument and the conformal value \(\xi=1/6\)}
As demonstrated in \cite{faraoniConformallyCoupledInflation2013,sonegoCouplingCurvatureScalar1993}, the coefficient appearing in Eq.~\eqref{eq:W} controls whether the leading short-distance tail vanishes.  If, at a spacetime point $x$, the quantity
\begin{equation}\label{eq:cancellation}
\mu^2 + \Big(\xi-\tfrac{1}{6}\Big)R(x)
\end{equation}
vanishes, then the leading contribution to the tail coefficient $W$ is absent at $x$.  Consequently, a \emph{massive} scalar would, at that location, propagate \emph{strictly} on the light cone, a manifestly unphysical behaviour because massive excitations are expected to propagate inside the light cone. As pointed out by \cite{faraoniConformallyCoupledInflation2013}, one may even arrange a constant-curvature background for which the curvature contribution cancels the mass contribution whenever $\xi\neq 1/6$.  Hence, the choice $\xi=0$ (minimal coupling) admits the possibility of such a cancellation and the attendant causal pathology.

The only way to preclude this pathology generically is to choose
\[
\xi=\frac{1}{6},
\]
for which the curvature-dependent piece in Eq.~\eqref{eq:W} vanishes identically in the coincidence limit.  For $\xi=1/6$ curvature cannot cancel the mass term and massive modes retain the expected timelike propagation.  Importantly, this conclusion follows solely from local properties of the retarded Green's function and the light-cone structure; conformal invariance is \emph{not} assumed a priori but emerges as the distinguished condition that enforces causal consistency in the general case. Non-minimal coupling plays a central role in a range of physical applications. Considering the form of the action \eqref{eq:action}, we see that non-minimal coupling introduces an effective squared mass in curved backgrounds given by
\[
\mu_{\mathrm{eff}}^2 = \mu^2 + \xi R.
\]
This effective squared mass term determines particle production in expanding cosmologies \cite{calzaImportanceBeingNonminimally2025}, the structure of late-time tails and quasinormal mode spectra in black-hole spacetimes \cite{santosQuasinormalModesMassive2021}, and features of early-universe dynamics \cite{faraoniConformallyCoupledInflation2013}. For these reasons, one should include $\xi$ as a free parameter in any analysis of scalar fields in curved spacetime and fix it either by symmetry principles, phenomenological requirements, or renormalization conditions.  If the absence of the causal pathology discussed above is imposed as a fundamental requirement, the conformal value $\xi=1/6$ is singled out in four dimensions.

\section{A Conformally Coupled Scalar Superradiance Problem}\label{sect:ACCSSP}
We now consider the behavior of a conformally-coupled scalar field, $\Phi$, of rest mass $\mu$ and charge $q$ in both spacetimes under consideration. The equation governing a massive, charged, conformally-coupled scalar field $\Phi$ in 4D is given by:

\beq \label{KGEq}
\left(D^{\nu}D_{\nu}-\mu^2-\frac{1}{6}R\right)\Phi=0, \text{ where } D_{\nu}=\nabla_{\nu}-iqA_{\nu}.
\eeq

Since the background spacetime is axial-symmetric and stationary, we use the following natural ansatz
\beq \label{ansatz}
\Phi=e^{-i\omega t}e^{im\phi}S(\theta)R(r),
\eeq
which allows us to separate angular and radial degrees of freedom to obtain
\beq \label{RadialEq}
\frac{d}{dr}\left(\Delta_r \frac{dR_{lm}(r)}{dr}\right)+\left(\frac{\Xi^2 \left(K(r)-\frac{qQr}{\Xi}\right)^2}{\Delta_r}-\mueff^2 r^2+\frac{\sigma Q^2 r}{6M}-\lambda_{lm}\right)R_{lm}(r)=0,
\eeq
\beq \label{AngularEq}
\frac{1}{\sin \theta}\frac{d}{d\theta}\left(\Delta_{\theta} \sin \theta \frac{dS_{lm}(\theta)}{d\theta}\right)-\left(\frac{\Xi^2\left(m-a\omega \sin^2(\theta)\right)^2}{\sin^2(\theta)\Delta_{\theta}}+a^2\mueff^2 \cos^2(\theta)-\lambda_{lm}\right)S_{lm}(\theta)=0.
\eeq
Here, $K(r)=\omega(r^2+a^2)-am$, $\mueff^2=\mu^2+\frac{2}{3}\Lambda$, and $\lambda_{lm}$ are the spheroidal eigenvalues of the angular operator given in \eqref{AngularEq}. Additionally, we include an indicator parameter, $\sigma$, in the above equations, 
\beq \label{sigma}
\sigma = \begin{cases}
0, & \text{for KNdS} \\
1, & \text{for KNdSCG}
\end{cases}
\eeq
which enables us to switch directly between the $KNdS$ and $KNdSCG$ cases.

As discussed in Sec.~II, the function $\Delta_r$ differs between the two metrics only in the coefficients of its cubic and constant terms. In both cases, 
$\Delta_r$ is a quartic polynomial with an identical leading-order term. For $\Lambda>0$, its roots may be ordered as $r_n<0<r_-<r_+<r_c$. Exploiting this common structure, we write $\Delta_r(r)=-\frac{\Lambda}{3}(r-r_+)(r-r_-)(r-r_c)(r-r_n)$ for both spacetimes. In this parametrization, the differences between the $KNdS$ and $KNdSCG$ solutions are entirely encoded in the specific values of the horizon radii and in the additional $\frac{\sigma Q^2 r}{6M}$ contribution.

We also note that the indicator parameter $\sigma$ does not appear in the angular equation \eqref{AngularEq}. Since the two metrics differ only through $\Delta_r$, it follows that the angular dependence of the conformally coupled scalar field governed by \eqref{KGEq} is identical in both spacetimes. For $\Lambda \ll 1$ and $a\omega\ll 1$, the angular equation becomes the equation for the spherical harmonics, resulting in eigenvalues $\lambda_{lm} = l(l+1)$. By defining the function $\psi(r)=\sqrt{r^2+a^2}R_{lm}(r)$ and tortoise coordinate $dr_{*}=\frac{r^2+a^2}{\Delta_r}dr$, the radial equation \eqref{RadialEq} takes the form \cite{britoSuperradiance2020Edition2020}
\beq \label{SEform}
\frac{d^2\psi(r_*)}{dr_{*}^2}+V(r)\psi(r_{*})=0,
\eeq
with the effective potential $V(r)$ given by
\beq \label{Potential}
V(r)=\frac{\Xi^2\left(K(r)-\frac{qQr}{\Xi}\right)^2-\Delta_{r}\left(r^2\mueff^2-\frac{\sigma Q^2 r}{6M}+\lambda_{lm}\right)}{\left(r^2+a^2\right)^2}-\frac{r\Delta_{r}\Delta_{r}^{'}}{\left(
r^2+a^2\right)^3}-\frac{\Delta_{r}^2\left(a^2-2r^2\right)}{\left(r^2+a^2\right)^4}.
\eeq
To study the scattering of the scalar field, we first evaluate the asymptotic values of the effective potential \eqref{Potential} at the outer ($r_+$) and cosmological horizons ($r_c$)
\beq \label{VatHor}
\begin{split}
& V(r_+)=\left(\Xi \omega-q\Phi_{+}-m\Omega_{+}\right)^2\equiv k_{+}^2, \text{ where } \Phi_{+}\equiv \frac{Qr_{+}}{r_{+}^2+a^2} \text{ and } \Omega_{+}\equiv \frac{\Xi a}{r_{+}^2+a^2}, \\
& V(r_c) = \left(\Xi \omega-q\Phi_{c}-m\Omega_{c}\right)^2\equiv k_{c}^2, \text{ where } \Phi_{c}\equiv \frac{Qr_{c}}{r_{c}^2+a^2} \text{ and } \Omega_{c}\equiv \frac{\Xi a}{r_{c}^2+a^2},
\end{split}
\eeq
where $\Omega_{+}$ and $\Omega_{c}$ are the angular velocities of the outer and cosmological horizons, respectively. The solutions to \eqref{SEform} will therefore approach plane waves in tortoise coordinate $r_{*}$ as $r\to r_+$ and as $r\to r_c$. Imposing solely ingoing waves as $r\to r_+$ yields the boundary conditions
\beq \label{BoundCond1}
\begin{split}
& \psi(r_*)\to Te^{-ik_+r_*}, \text{ for } r\to r_+,\\
& \psi(r_*)\to I e^{-ik_cr_*}+R e^{ik_cr_*}, \text{ for } r\to r_c,
\end{split}
\eeq
where $T$ is the ingoing wave amplitude as $r\to r_+$, and $I$ and $R$ are the ingoing and outgoing wave amplitudes as $r\to r_c$, respectively. Computing and equating the Wronskian quantity of $\psi$ and its complex conjugate $\psi^{*}$ at both boundaries 
\beq \label{Wrons}
W=(U\frac{dU^*}{dr_*}-U^*\frac{dU}{dr_*}),
\eeq
we obtain the following condition on the squared moduli of the boundary amplitudes 
\beq \label{ConEq}
|R|^2=|I|^2-\frac{k_{+}}{k_{c}}|T|^2.
\eeq
For superradiance to occur, we require that the amplitude of the reflected wave be larger than the amplitude of the incident wave $(|R|>|I|$, i.e. $\frac{k_+}{k_c} < 0$ in the above equation). This leads to the superradiance condition
\beq \label{SRCond}
q\Phi_c+m\Omega_c < \Xi\omega < q\Phi_+ + m\Omega_+.
\eeq
We note that this condition implies that the functional form of the superradiant frequency range is identical for both the $KNdS$ and $KNdSCG$ spacetimes. However, because the horizon radii $r_+$ and $r_c$ depend on the chosen metric through $\Delta(r)$, the frequency ranges change accordingly.

In addition to the superradiant frequency range, we are also interested in quantifying the magnitude of superradiant amplification, given by the amplification factor,
\beq \label{AmpFacDef}
Z_{lm} = \frac{|R|^2}{|I|^2}-1.
\eeq
The evaluation of $Z_{lm}$ requires solving the radial equation \eqref{RadialEq}. As this equation is not easily solvable in analytical form, we now split our analysis into two subcases: the case of a massless scalar field ($\mu=0$) and that of a massive scalar field ($\mu \ne0$). For each case, we will use different techniques to analyze the behavior of superradiance phenomena.

\section{Superradiance of a Massless Scalar Field}\label{sect:SoaMAF}
In this section, we study the superradiance of a massless, charged, conformally-coupled scalar in the $KNdS$ and $KNdSCG$ spacetimes. As $\mu=0$, the effective mass becomes $\mueff^2=\frac{2}{3}\Lambda$. Assuming the black hole is not extremal, the radial equation \eqr{RadialEq} has exactly five regular singular points on the Riemann sphere: four finite regular singular points at $\Delta_r(r)=0$ given by $r_n<0<r_-<r_+<r_c$, and one regular singular point at $\infty$. When $\mu=0$, the singular point at $z=u$ $(r=r_n)$ is removable. Similar to the previous work for the $KNdS$ background in \cite{suzukiPerturbationsKerrdeSitter1998,baticHeunEquationTeukolsky2007}, we make the following coordinate transformation
\beq \label{zCoord}
z=\left(\frac{r_c-r_n}{r_c-r_-}\right)\left(\frac{r-r_-}{r-r_n}\right)\equiv u\left(\frac{r-r_-}{r-r_n}\right).
\eeq
Additionally, we define a new function, $H_{lm}(z)$, such that
\[
R_{lm}(z)=z^{\rho_1}(z-1)^{\rho_2}(z-w)^{\rho_3}(z-u)^1H_{lm}(z),
\]
where 
\[ w\equiv u\left(\frac{r_+-r_-}{r_+-r_n}\right),
\]
\[
 \rho_1 = \frac{3 i \Xi \left(K(r_-)-\frac{qQr_-}{\Xi}\right)}{\Lambda (r_c-r_-)(r_n-r_-)(r_+-r_-)},\quad \rho_2 = -\frac{3 i \Xi \left(K(r_c)-\frac{qQr_c}{\Xi}\right)}{\Lambda (r_--r_c)(r_n-r_c)(r_+-r_c)},
\]
\[
\rho_3 = \frac{3 i \Xi \left(K(r_+)-\frac{qQr_+}{\Xi}\right)}{\Lambda (r_c-r_+)(r_--r_+)(r_n-r_+)},\quad \rho_4 = \frac{3 i \Xi \left(K(r_n)-\frac{qQr_n}{\Xi}\right)}{\Lambda (r_c-r_n)(r_--r_n)(r_+-r_n)}.
\]
In terms of $H_{lm}(z)$ , the radial equation becomes
\beq \label{HeunEq}
\frac{d^2 H_{lm}}{dz^2}+\left(\frac{\gamma}{z}+\frac{\delta}{z-1}+\frac{\epsilon}{z-w}\right)\frac{dH_{lm}}{dz}+\left(\frac{\alpha \beta z - k}{z(z-1)(z-w)}\right)H_{lm} = 0,
\eeq
where
\[
\alpha= 1-\rho_4+\sum_{i=1}^{3} \rho_i, \quad \beta= 1+\rho_4+\sum_{i=1}^{3} \rho_i, \quad \gamma = 2\rho_1+1, \quad\delta = 2\rho_2+1, \quad\epsilon = 2\rho_3+1,
\]
\[
\begin{split}
k = & \frac{1}{\left(r_+-r_n\right)\left(r_c-r_-\right)\left(r_+-r_-\right){}^2} \Biggl(\frac{\sigma Q^2 r_- \left(r_+-r_-\right){}^2}{2M\Lambda}+\frac{1}{\Lambda^2 \left(r_c-r_-\right){}^2 \left(r_--r_n\right)}\biggl(-2 \Lambda ^2 r_-^7 \\
&-18 a^2 m^2 \Xi ^2 r_c+18 \Xi  q Q \bigl(a^2 \omega  ((r_-+r_+) r_c+r_- (r_+-3 r_-))-a m ((r_-+r_+) r_c+r_- (r_+-3 r_-))\\
&-r_-^2 \omega  ((r_--3 r_+) r_c+r_- (r_-+r_+))\bigr)+36 a m \Xi ^2 \omega  \bigl(a^2 (r_c-2 r_-+r_+)+r_+ r_- r_c-r_-^3\bigr)\\
&+18 \Xi ^2 \omega ^2 (a^2+r_-^2) \bigl(r_- (-2 r_+ r_c+r_- r_c+r_- r_+)-a^2 (r_c-2 r_-+r_+)\bigr)-18 a^2 m^2 \Xi ^2 r_+\\
&+2 \Lambda  (r_--r_+){}^2 r_- r_c (r_--r_n) (3 \lambda _{lm}+2 \Lambda  r_-^2)-\Lambda  (r_--r_+){}^2 r_c^2 (r_--r_n) \bigl(3 \lambda _{lm}+2 \Lambda  r_-^2\bigr)\\
&+18 q^2 Q^2 r_- (r_-^2-r_+ r_c)+3 \Lambda  r_-^4 \lambda _{lm} r_n-6 \Lambda  r_+ r_-^3 \lambda _{lm} r_n+3 \Lambda  r_+^2 r_-^2 \lambda _{lm} r_n-3 \Lambda  r_-^5 \lambda _{lm}\\
&+6 \Lambda  r_+ r_-^4 \lambda _{lm}-3 \Lambda  r_+^2 r_-^3 \lambda _{lm}+2 \Lambda ^2 r_-^6 r_n-4 \Lambda ^2 r_+ r_-^5 r_n+2 \Lambda ^2 r_+^2 r_-^4 r_n+4 \Lambda ^2 r_+ r_-^6-2 \Lambda ^2 r_+^2 r_-^5\biggr)\Biggr)\\
&+w\left(\gamma \rho_2 +\rho_1+\frac{1}{u}\right)+\left(\gamma \rho_3 +\rho_1 \right).
\end{split}
\]
Equation \eqr{HeunEq} is the most general second-order linear differential equation with four regular singular points, called the General Heun Equation.\footnote{For a review of Heun-like equations applications in physics, we direct the reader to \cite{hortacsuHeunFunctionsTheir2018}.} Local solutions to Heun's equation around $z=w$ $(r\to r_{+})$ and $z=1$ $(r\to r_c)$ are expressible in terms of the local Heun function $H\ell (w,k;\alpha, \beta, \gamma, \delta; z)$ \cite{DLMFChapter31}. The local Heun function is defined such that
\small
\beq \label{LocalHeun}
H\ell (w,k;\alpha, \beta, \gamma, \delta; z) = 1+\frac{k}{\gamma w}z + \frac{k^2+k(1+\alpha+\beta-\delta+w(\gamma+\delta))-\alpha \beta \gamma w}{2\gamma (\gamma +1) w^2}z^2 + O(z^3).
\eeq
\normalsize
In terms of $H\ell$, the two linearly independent solutions of \eqref{HeunEq} for $z\sim w$ $(r\sim r_+)$ are
\beq \label{Heunatw}
\begin{aligned}
&h_-^{(w)} (z) =H\ell \biggl(\frac{w}{w-1},\frac{k-w\alpha\beta}{1-w},\alpha,\beta,\epsilon,\delta,\frac{z-w}{1-w}\biggr), \\ 
&h_+^{(w)} (z) =(z-w)^{1-\epsilon}H\ell \biggl(\frac{w}{w-1},\frac{k-(\beta-\gamma-\delta)(\alpha-\gamma-\delta)w-\gamma(\epsilon-1)}{1-w},\\
&\ \ \ \ \ \ \ \ \ \ \ \ \ \ \ \ \ \ \ \ \ \ \ \ \ \ \ \ \ \ \ \ \ \  -\alpha+\gamma+\delta,-\beta+\gamma+\delta,2-\epsilon,\delta,\frac{z-w}{1-w}\biggr),
\end{aligned}
\eeq
and the ones for $z\sim 1$ $(r\sim r_c)$ are
\beq \label{Heunat1}
\begin{aligned}
h_-^{(1)}(z)=&\biggl(\frac{z-w}{1-w}\biggr)^{-\alpha}H\ell \biggl(w,k+\alpha(\delta-\beta),\alpha,\delta+\gamma-\beta,\delta,\gamma,w\frac{1-z}{w-z}\biggr),\\
h_+^{(1)}(z)=&(z-1)^{1-\delta}\biggl(\frac{z-w}{1-w}\biggr)^{\delta-\alpha-1}H\ell \biggl(w,k-(\delta-1)\gamma w-(\beta-1)(\alpha-\delta+1),\\
&\ \ \ \ \ \ \ \ \ \ \ \ \ \ \ \ \ \ \ \ \ \ \ \ \ \ \ \ \ \ \ \ \ \  -\beta+\gamma+1,\alpha-\delta+1,2-\delta,\gamma,w\frac{1-z}{w-z}\biggr).
\end{aligned}
\eeq
In terms of these local series solutions, the general solution to the radial equation in the massless case is given by
\beq \label{RadialEqSol}
R_{lm}(z)=z^{\rho_1}(z-1)^{\rho_2}(z-w)^{\rho_3}(z-u)\left(c_1 h_+^{(w)}(z) + c_2 h_-^{(w)}(z) \right)
\eeq
for some $c_1,c_2\in \C$. Next, we discuss our approach to further reduce the above equation in the two limiting cases of $z\to w$ and $z\to 1$. 

Taking the limit of $z\to w$ in Eqs.~\eqr{Heunatw} and \eqr{LocalHeun}, the above equation becomes 
\[
R_{lm}(z)\sim w^{\rho_1}(w-1)^{\rho_2}(w-u)\left(c_1 (z-w)^{-\rho_3} + c_2 (z-w)^{\rho_3} \right), \text{ as } z\to w.
\]
In terms of tortoise coordinate, $dr_{*}=\frac{r^2+a^2}{\Delta_r}dr$, we have
\[
\left(z-w\right)^{\pm \rho_3} \sim \left(\frac{(r_c-r_n)(r_--r_n)}{(r_c-r_-)(r_+-r_n)}\right)^{\pm \rho_3} \cdot e^{\pm i k_{+} r_{*}}, \text{ as } z\to w.
\]
The coefficient $c_2$ vanishes by imposing boundary conditions \eqr{BoundCond1}. As a consequence of the linearity of \eqr{HeunEq}, it must be possible to write $h_+^{(w)}(z)$ (or its analytic continuation) as a linear combination of $h_-^{(1)}(z)$ and $h_+^{(1)}(z)$,
\[
h_+^{(w)}(z) = C_- h_-^{(1)}(z) + C_+h_+^{(1)}(z).
\]
Plugging this connection formula into \eqr{RadialEqSol}, and taking the limit of $z\to 1$ in Eqs.~\eqr{Heunatw} and \eqr{LocalHeun}, we can reduce Eq.~\eqr{RadialEqSol} to
\[
R_{lm}(z)\sim (1-w)^{\rho_3}(1-u)\left(c_1C_- (z-1)^{\rho_2} + c_1C_+ (z-1)^{-\rho_2} \right), \text{ as } z\to 1.
\]

Also, note the $\left(z-1\right)^{\pm \rho_2}$ terms can be rewritten in terms of horizons
\[
\left(z-1\right)^{\pm \rho_2} \sim \left(\frac{r_--r_n}{(r_c-r_-)(r_+-r_n)}\right)^{\pm \rho_2} \cdot e^{\mp i k_c r_{*}}, \text{ as } z\to 1.
\]
Imposing the boundary condition \eqr{BoundCond1}, we arrive at
\beq \label{Amplitudes}
\begin{split}
& T=c_1(r_+^2+a^2)^{1/2}w^{\rho_1}(w-1)^{\rho_2}(w-u)\left(\frac{(r_c-r_n)(r_--r_n)}{(r_c-r_-)(r_+-r_n)}\right)^{-\rho_3},\\
& R=c_1(r_c^2+a^2)^{1/2}(1-w)^{\rho_3}(1-u)\left(\frac{r_--r_n}{(r_c-r_-)(r_+-r_n)}\right)^{-\rho_2} C_{+},\\
& I=c_1(r_c^2+a^2)^{1/2}(1-w)^{\rho_3}(1-u)\left(\frac{r_--r_n}{(r_c-r_-)(r_+-r_n)}\right)^{\rho_2} C_{-}.
\end{split}
\eeq
The above amplitude equations \eqr{Amplitudes} can then be used to compute the amplification factor $Z_{lm}$ defined in \eqr{AmpFacDef} 
\beq \label{AmpFac2}
\begin{split}
Z_{lm} = \frac{|R|^2}{|I|^2}-1&=\left|\left(\frac{r_--r_n}{(r_c-r_-)(r_+-r_n)}\right)^{-2\rho_2}\right|\cdot\left|\frac{C_+}{C_-}\right|^2-1.
\end{split}
\eeq
Note that from the horizon ordering $(r_n<0<r_-<r_+<r_c)$, we have
\[
\frac{r_--r_n}{(r_c-r_-)(r_+-r_n)} > 0.
\]
From the definition of $\rho_2$, it follows that $\Re(\rho_2)=0$. Thus, we can rewrite
\[
\left(\frac{r_--r_n}{(r_c-r_-)(r_+-r_n)}\right)^{-2\rho_2}=\exp\left(-2i\Im(\rho_2)\ln\left(\frac{r_--r_n}{(r_c-r_-)(r_+-r_n)}\right)\right),
\]
which has a modulus of 1. Thus, $Z_{lm}$ can be simplied to
\beq \label{AmpFac3}
Z_{lm} = \frac{|C_+|^2}{|C_-|^2}-1.
\eeq

It is clear in \eqr{AmpFac3} that the problem of computing the amplification spectrum of either black hole is mathematically equivalent to solving for the connection coefficients $C_+,C_-$ as functions of the Heun parameters $\{\alpha,\beta,\gamma,\delta,\epsilon,k\}$. The Heun connection problem is a rich and analytically complex subject in the mathematics of differential equations. There exist several techniques to obtain $C_+, C_-$ perturbatively for certain regimes of $|w|$ \cite{hortacsuHeunFunctionsTheir2018,bonelliIrregularLiouvilleCorrelators2023,aminovBlackHolePerturbation2023,lisovyyPerturbativeConnectionFormulas2022}.  For the purposes of this paper, we will use the Heun-CFT correspondence approach developed by Bonelli et al. \cite{bonelliIrregularLiouvilleCorrelators2023}, where it was demonstrated the Heun connection problem could be solved perturbatively in $0<|w|\ll 1$ by matching the Heun equation to the semiclassical limit of the corresponding BPZ equation satisfied by a five-point correlation function with one degenerate insertion into a Liouville type CFT. Using the same dictionary convention defined in \cite{aminovBlackHolePerturbation2023}, we obtain
\beq \label{dict}
\begin{split}
& a_0 = \frac{1-\gamma}{2}=-\rho_1,\\
& a_1 = \frac{1-\delta}{2}=-\rho_2,\\
& a_w = \frac{1-\epsilon}{2}=-\rho_3,\\
& a_{\infty} = \frac{\alpha-\beta}{2}=-\rho_4,\\
& \E^{(0)} = \frac{2w\alpha\beta+\gamma\epsilon-w(\gamma+\delta)\epsilon-2k}{2(w-1)},
\end{split}
\eeq
where $\{a_0,a_1,a_w,a_{\infty}\}$ are the semiclassical momenta and $\E^{(0)}$ is the leading acessory parameter of the associated CFT in the semiclassical limit. $\E^{(0)}$ is related to the classical conformal block $F$, semiclassical momenta $\{a_0,a_1,a_w,a_{\infty}\}$, and the intermediate Liouville momentum $\aint$, by the Matone relation
\[
\E^{(0)}=-\frac{1}{4}-\aint^2+a_w^2+a_0^2+w\partial_w F(w).
\]
The classical conformal block, $F$, can be combinatorially expanded in $w$ via AGT correspondence as described in \cite{bonelliIrregularLiouvilleCorrelators2023},
\beq \label{Fexpansion}
F(w) = \frac{\left(\frac{1}{4}-\aint^2-a_1^2+a_{\infty}^2\right)\left(\frac{1}{4}-\aint^2-a_w^2+a_{0}^2\right)}{\left(\frac{1}{2}-2\aint^2\right)}w+O(w^2).
\eeq
Plugging \eqr{Fexpansion} into the Matone relation, we invert order by order in $w$ to obtain an expansion for $\aint$ as
\small
\[
\aint=\pm \sqrt{-\frac{1}{4}-\E^{(0)}+a_w^2+a_0^2}\left(1-\frac{\left(2a_0^2+2a_1^2+2a_w^2-2a_{\infty}^2-2\E^{(0)}-1\right)\left(4 a_w^2-2\E^{(0)}-1\right)}{2\left(4 a_0^2+4a_w^2-4\E^{(0)}-1\right)\left(2a_0^2+2a_w^2-2\E^{(0)}-1\right)}w+O(w^2)\right).
\]
\normalsize
Defining the same hypergeometric connection matrix as in \cite{aminovBlackHolePerturbation2023},
\[
\mathcal{M}_{\theta \theta'}(a_1,a_2;a_3) = \frac{\Gamma(-2\theta'a_2)\Gamma(1+2\theta a_1)}{\Gamma\left(\frac{1}{2}+\theta a_1-\theta' a_2 + a_3\right) \Gamma\left(\frac{1}{2}+\theta a_1-\theta' a_2 - a_3\right)},
\] 
the connection formula between $z\sim w$ and $z\sim 1$ is then given by \cite{bonelliIrregularLiouvilleCorrelators2023}
\[
\begin{aligned}
&w^{-\frac{1}{2}+a_0-a_w}(1-w)^{-\frac{1}{2}+a_1}e^{-\frac{1}{2}\partial_{a_w}F(w)}h_+^{(w)}(z)=\\
&\left(\sum_{\nu=\pm}\mathcal{M}_{+\nu}(a_w,\aint;a_0)\mathcal{M}_{(-\nu)-}(\aint,a_1;a_{\infty})w^{\nu \aint}e^{-\frac{\nu}{2}\partial_{\aint}F(w)}\right)(1-w)^{-\frac{1}{2}+a_w}e^{i\pi(a_1+a_w)}e^{\frac{1}{2}\partial_{a_1}F(w)}h_-^{(1)}(z)+\\
&\left(\sum_{\nu=\pm}\mathcal{M}_{+\nu}(a_w,\aint;a_0)\mathcal{M}_{(-\nu)+}(\aint,a_1;a_{\infty})w^{\nu \aint}e^{-\frac{\nu}{2}\partial_{\aint}F(w)}\right)(1-w)^{-\frac{1}{2}+a_w}e^{i\pi(-a_1+a_w)}e^{-\frac{1}{2}\partial_{a_1}F(w)}h_+^{(1)}(z).
\end{aligned}
\]
This leads us to obtain the ratio of connection coefficients as
\beq \label{Ratio}
\frac{C_+}{C_-} = \left(\frac{\sum_{\nu=\pm}\mathcal{M}_{+\nu}(a_w,\aint;a_0)\mathcal{M}_{(-\nu)+}(\aint,a_1;a_{\infty})w^{\nu \aint}e^{-\frac{\nu}{2}\partial_{\aint}F(w)}}{\sum_{\nu=\pm}\mathcal{M}_{+\nu}(a_w,\aint;a_0)\mathcal{M}_{(-\nu)-}(\aint,a_1;a_{\infty})w^{\nu \aint}e^{-\frac{\nu}{2}\partial_{\aint}F(w)}}\right) e^{-2i\pi a_1} e^{-\partial_{a_1}F(w)},
\eeq
where the necessary derivatives of $F(w)$ can be obtained using \eqr{Fexpansion} and the corresponding expansion for $\aint$
\[
\partial_{\aint} F(w)=\frac{\aint \left(-16 \aint^4+8 \aint^2+16 \left(a_w^2-a_0^2\right) (a_1-a_{\infty}) (a_1+a_{\infty})-1\right)}{\left(1-4 \aint^2\right)^2}w+O\left(w^2\right)
\]
\[
\partial_{a_1} F(w) =\frac{a_1\left(-4 \aint^2+4 a_0^2-4 a_w^2+1\right)}{4 \aint^2-1}w+O\left(w^2\right)
\]

We have therefore solved for the amplification spectrum perturbatively in $0<w\ll 1$. Recall that the crossing ratio, $w$, is defined by
\[
w = \frac{r_c-r_n}{r_c-r_-}\frac{r_+-r_-}{r_+-r_n}.
\]
Consider the value of $w$ for $\Lambda \ll 1$. In the $KNdS$ metric, this corresponds to $r_n \sim -r_c$ and thus
\[
w^{(KNdS)}\sim \frac{2 r_c}{r_c}\frac{r_+-r_-}{r_c} = \frac{2(r_+-r_-)}{r_c} \ll 1.
\]
We therefore conclude that $w^{(KNdS)}\ll 1$ is true independent of any black hole parameter for $\Lambda \ll 1$. This makes the perturbative approach particularly accurate over nearly all allowed parameter values for realistic $\Lambda \ll 1$. Alternatively, in the $KNdSCG$ metric, we have $r_n \sim -\frac{6M}{Q^2}$ (see \ref{HorizonApprox}). This results in
\[
w^{(KNdSCG)}\sim \frac{r_c}{r_c}\frac{r_+-r_-}{r_+-r_n} \sim \frac{Q^2(r_+-r_-)}{6M} \ll 1.
\]
Unlike $KNdS$, $w^{(KNdSCG)}\ll 1$ requires small $Q^2$ to be accurate in the $\Lambda \ll 1$ regime. Alternatively, note the factors of $(r_+-r_-)$ in both $w^{(KNdSCG)}$ and $w^{(KNdS)}$. We therefore expect $w^{(KNdSCG)}\to w^{(KNdS)}\to 0$ in the extremal limit $r_+\to r_-$. We conclude that our perturbative approach is expected to yield accurate results for both metrics in the $0<\Lambda \ll Q \ll 1$ regime and/or the extremal limit. In the $KNdS$ background, the condition that $Q^2\ll 1$ is not necessary and the perturbative approach is expected to be accurate generally for both $0<\Lambda \ll 1$ and the extremal limit. From this point onward, we will assume that $0<\Lambda \ll Q^2 \ll 1$ so that the approximate connection formula holds perturbatively for both metrics.

\begin{figure}[!htbp]
\includegraphics[width=1\linewidth, height=6cm]{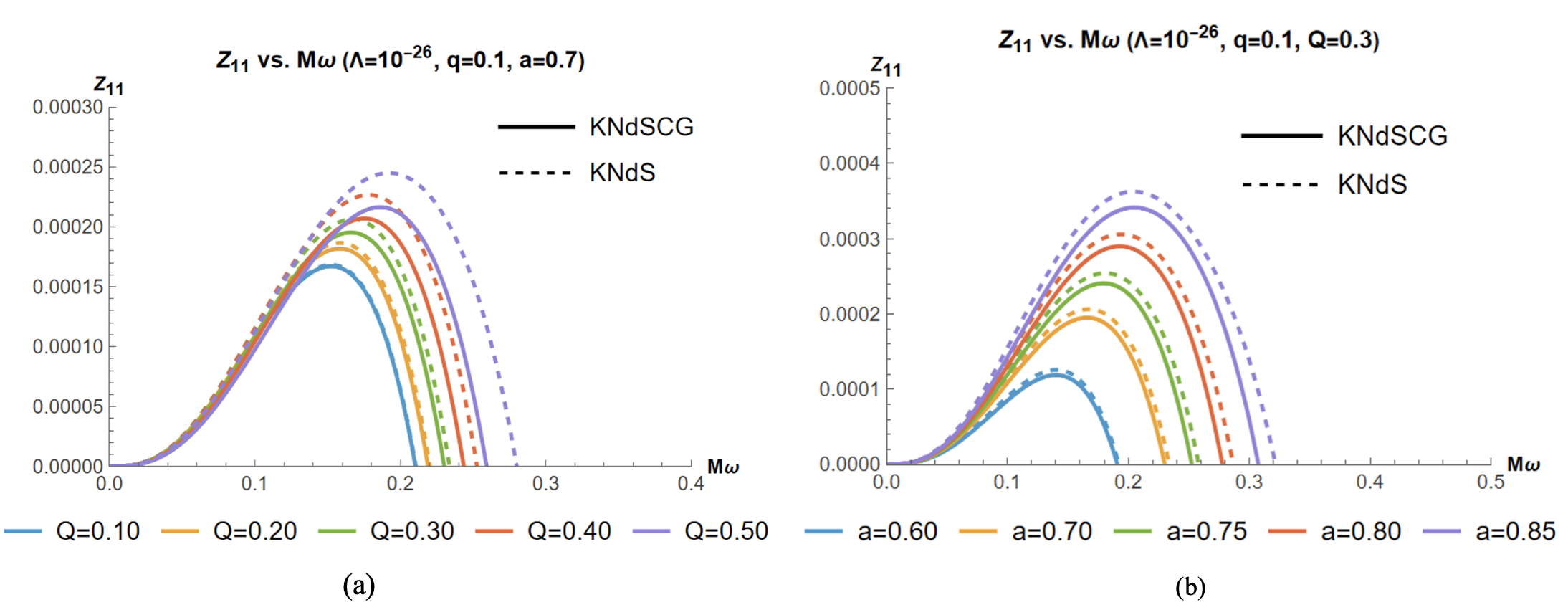} 
\caption{Amplification factor for the $l=1, m=1$ mode $Z_{11}$ as a function of scalar field frequency $M\omega$ for varying black hole charge $Q$ (Panel (a)) and spin $a$ (Panel (b)). In both graphs, the \textbf{solid lines} represent the $KNdSCG$ spacetime and the \textbf{dashed lines} represent the $KNdS$ spacetime.}
\label{Z11}
\end{figure}

\begin{figure}[!htbp]
\includegraphics[width=1\linewidth, height=6cm]{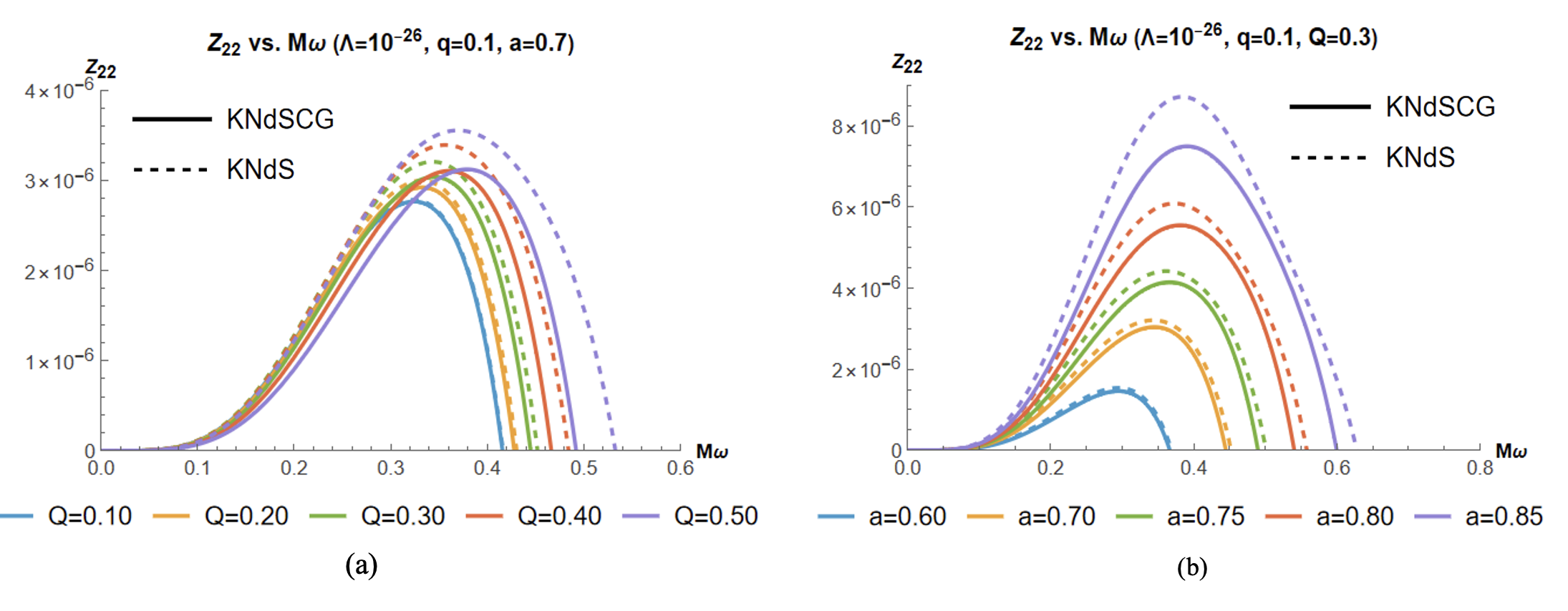} 
\caption{Amplification factor for the $l=2, m=2$ mode $Z_{22}$ as a function of scalar field frequency $M\omega$ for varying black hole charge $Q$ (Panel (a)) and spin $a$ (Panel (b)). In both graphs, the \textbf{solid lines} represent the $KNdSCG$ spacetime and the \textbf{dashed lines} represent the $KNdS$ spacetime.}
\label{Z22}
\end{figure}

\begin{figure}[!htbp]
\includegraphics[width=1\linewidth, height=6cm]{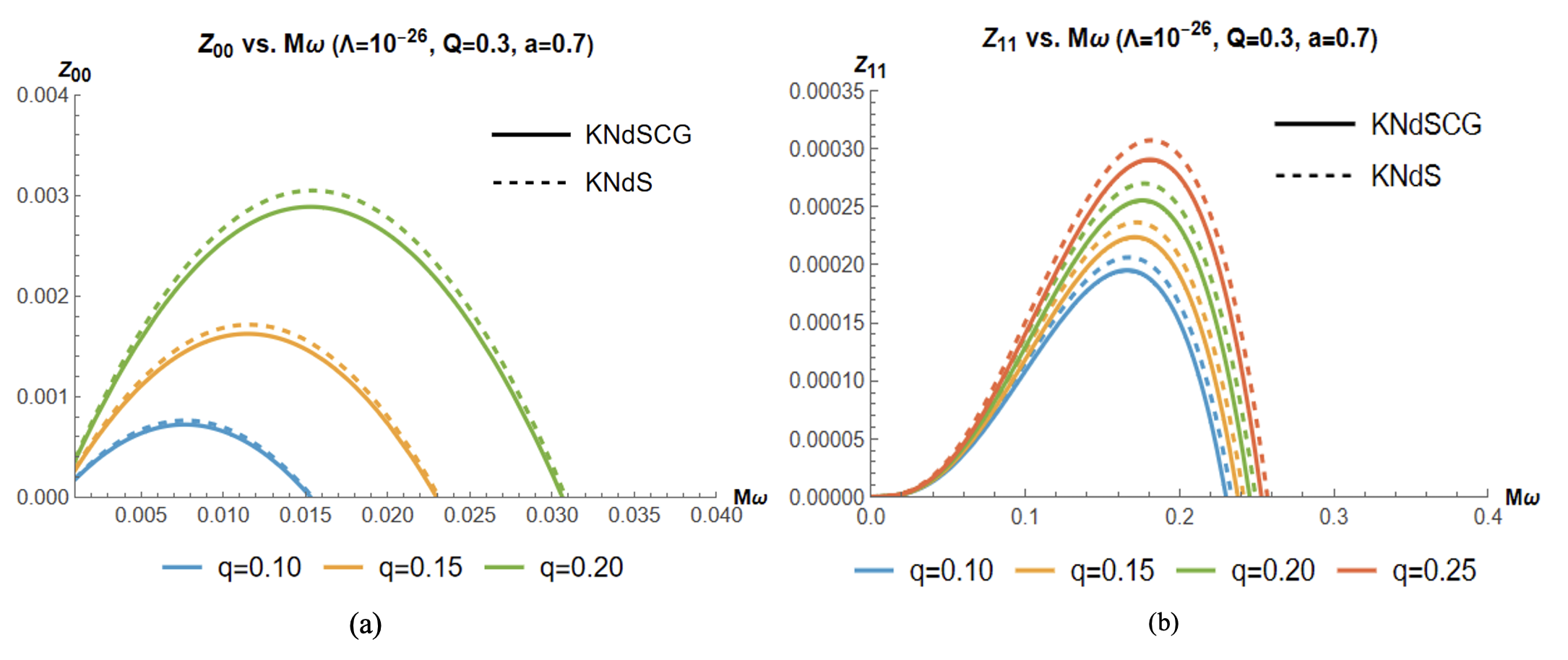} 
\caption{Amplification factors $Z_{00}$ (Panel (a)) and $Z_{11}$ (Panel(b)) for varying scalar field charge $q$. In both graphs, the \textbf{solid lines} represent the $KNdSCG$ spacetime and the \textbf{dashed lines} represent the $KNdS$ spacetime.}
\label{Varyq}
\end{figure}

Figs.~\ref{Z11} and \ref{Z22} show that both superradiant amplification factors $Z_{11}$ and $Z_{22}$ sourced by our conformal Weyl gravity metric \eqref{metric} are consistently suppressed compared to that sourced by the $KNdS$ metric. The peak frequency is also slightly deviated between the two spacetimes. As expected for both metrics, $Z_{11}$ and $Z_{22}$ increase monotonically with increasing $a$ and $Q$. The amplification spectra of the two black holes approach each other as $Q$ decreases until $Q=0 $ when both metrics reduce to the regular Kerr-dS metric.


\section{Superradiance of a Massive Scalar Field}\label{sect:SoaMiveSF}
The analysis of superradiance for a massive, charged, conformally coupled scalar field is more involved, as the radial equation cannot be reduced to Heun form: all five regular singular points remain irreducible. We therefore analyze this case using the WKB approximation. For the remainder of this section, we focus on the $KNdSCG$ background, and therefore treat the indicator as $\sigma=1$.

\subsection{The Turning Point of \texorpdfstring{$\mathbf{V(r)}$}{V(r)} and the Effective Far-Region}

\begin{figure}[ht]
\includegraphics[width=1\linewidth, height=6.5cm]{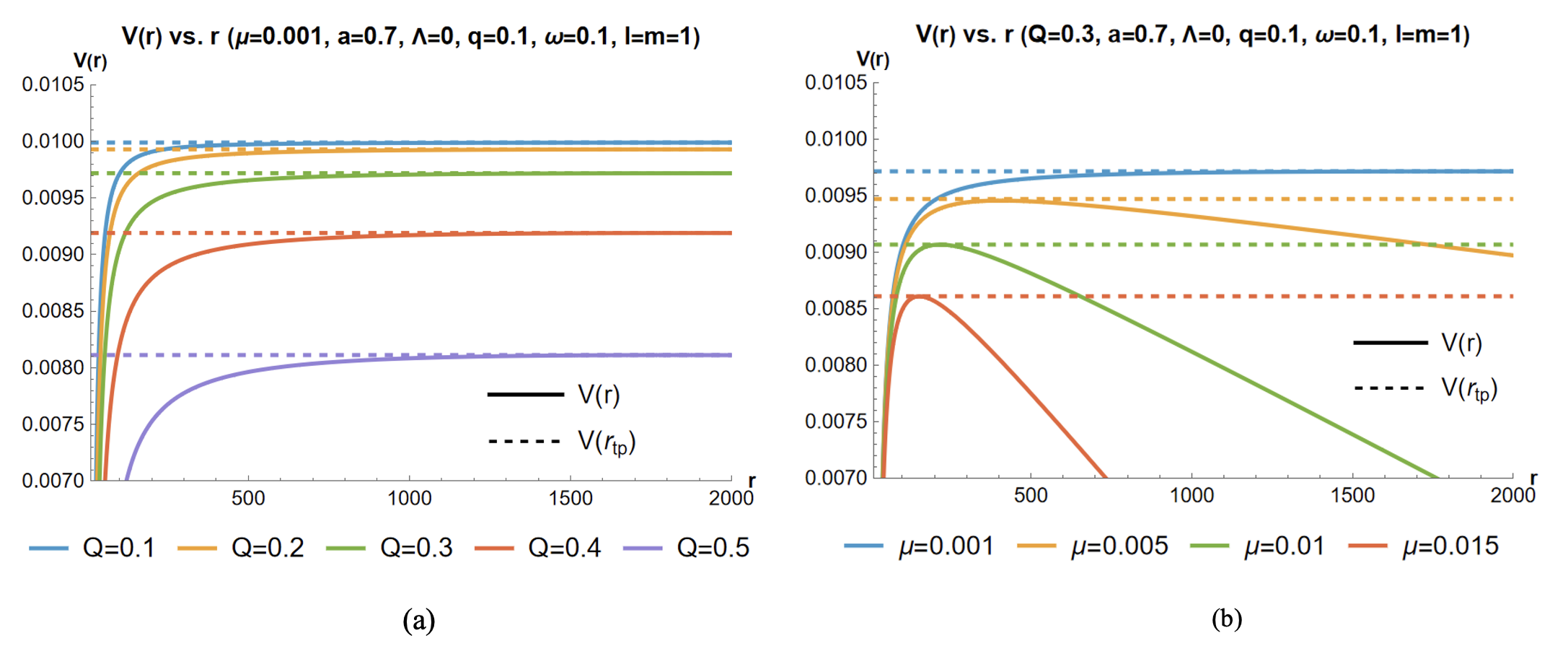} 
\caption{Effective potential $V(r)$ as a function of $r$ over $r\in [30,2000]$ for varying black hole charge $Q$ (Panel (a)) and scalar field mass $\mu$ (Panel (b)). $V(r_{tp})$ plotted in dashed line as a reference.}
\label{VFlat}
\end{figure}

\begin{figure}[ht]
\includegraphics[width=1\linewidth, height=6.5cm]{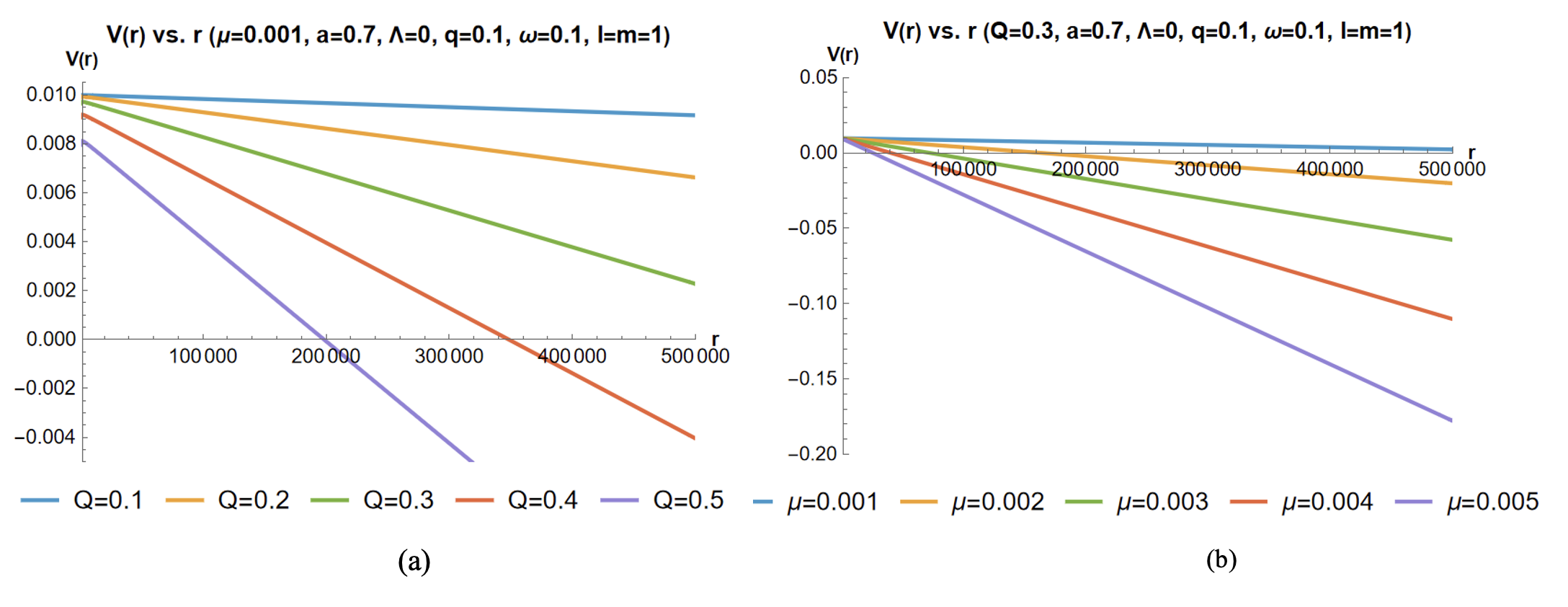} 
\caption{Effective potential $V(r)$ as a function of $r$ over $r\in [2000,500000]$ for varying blak hole charge $Q$ (Panel (a)) and scalar field $\mu$ (Panel (b)).}
\label{VLin}
\end{figure}

We begin by investigating the properties of $V(r)$ for $r_+\ll r\ll r_c$, assuming that $\Lambda \ll 1$ and thus $\Xi\approx 1$. The effective scalar potential given in Eq.~\eqref{Potential} then becomes,
\beq \label{PotentialNoL}
V(r)=\frac{\left(K(r)-qQr\right)^2-\Delta_{r}\left(r^2\mueff^2-\frac{Q^2 r}{6M}+\lambda_{lm}\right)}{\left(r^2+a^2\right)^2}-\frac{r\Delta_{r}\Delta_{r}^{'}}{\left(
r^2+a^2\right)^3}-\frac{\Delta_{r}^2\left(a^2-2r^2\right)}{\left(r^2+a^2\right)^4}.
\eeq
The behavior of the effective potential when $r\gg 1$ can be probed by series expanding it to the order of $O(r^{-1})$, yielding
\beq \label{VeffLarger}
V(r)=\frac{\Lambda}{3}  \mu ^2 r^2-\frac{\mu ^2 Q^2r}{6 M}+\left(\frac{\Lambda}{3}  (\lambda_{lm} -\Xi +1)+\Xi  \left(\Xi  \omega ^2-\mu ^2\right)\right)+O\left(r^{-1}\right).
\eeq
Note that, when $\Lambda = 0$ and $Q^2\mu^2 \neq 0$, the above equation becomes linear in $r$. From Appendix \eqref{rc}, we show that $r_c\sim O(Q^2 \Lambda^{-1})$. Thus, for the observable range of $r_{+}\ll r \ll r_{c}$, the effective potential takes the form
\beq \label{BigRPotNoL}
V(r)=-\frac{\mu ^2 Q^2r}{6 M}+ \left(\omega ^2-\mu ^2\right)+O(r^{-1}),
\eeq
which represents a linearly decreasing $V(r)$ at large $r$ with a slope of approximately $-Q^2 \mu^2/6 M$. Figs.~\ref{VFlat} and \ref{VLin} show how the effective potential $V(r)$ varies as a function of $r$ for different values of cosmological constant $\Lambda$ (Panel (a)) and mass of the scalar field $\mu$ (Panel (b)). To capture both the short- and long-range behavior, we divide the analysis into two regimes: $r\in [30,2\times 10^3]$ and $r\in [2\times 10^3,5\times 10^5]$. From Fig.~\ref{VFlat}, we see that $V(r)$ first increases until it reaches a local maximum, denoted by $r_{tp}$, after which $V(r)$ then decreases in an approximately linear manner with slope $\approx-Q^2 \mu^2/6 M$. Fig.~\ref{VFlat} also shows that $V(r)$ is approximately constant with a value of $V(r_{tp})$ at its local maximum when $r=r_{tp}$ and $r_{tp}\gg 1$ for small scalar field mass $\mu\sim O(0.001)$. The size of this ``flat-region" grows with decreasing mass of the scalar field $\mu$, as seen in Fig.~\ref{VFlat}(b).  Therefore, in the large $r$ and small $\Lambda (\ll 1)$ regime, it is a reasonable approximation to expand $V(r)$ to the order of $O(r^{-2})$ without loss of its essential characteristics. This yields
\beq \label{Orm2PotExpNoL}
V(r)= \left(\omega ^2-\mu ^2\right) - \frac{\mu ^2 Q^2r}{6 M}+\frac{-\frac{Q^2 \left(\lambda_{lm} -2 a^2 \mu ^2\right)}{6 M}+2 \mu ^2 M-2 q Q \omega }{r}+O(r^{-2}).
\eeq
Taking the derivative of \eqref{Orm2PotExpNoL} with respect to the tortoise coordinate $r_{*}$, we get
\beq \label{Orm3DPotExpNoL}
\frac{dV}{dr_{*}} =-\frac{\mu ^2 Q^4 r}{36 M^2}-\frac{\mu ^2 Q^2}{6 M}+\frac{Q^3 \left(-a^2 \mu ^2 Q+12 M q \omega +\lambda_{lm}  Q\right)}{36 M^2 r}+O(r^{-2}).
\eeq
By definition, $r_{tp}$ is the location of a local maximum in $V(r_*)$ and thus $\frac{dV(r_{tp})}{dr_{*}}=0$. Solving for the positive roots of Eq.~\eqref{Orm3DPotExpNoL}, we get 
\beq \label{rtp}
r_{tp} \approx \frac{\sqrt{Q^4 \left(\lambda_{lm} -a^2 \mu ^2\right)+9 \mu ^2 M^2+12 M q Q^3 \omega}}{\mu Q^2}-\frac{3M}{Q^2} + O(r_{tp}^{-2}).
\eeq
Plugging this root into \eqref{Orm2PotExpNoL}, we obtain $V(r_{tp})$
\beq \label{Vtp}
V(r_{tp}) \approx \left(\omega^2 - \mu^2\right) -\frac{\mu  \sqrt{Q^3 (12 M q \omega +\lambda_{lm}  Q)}}{3 M}\equiv k_{far}^{2},\text{ for } Q^2\mu^2\ll 1.
\eeq
From Eq.~\eqref{rtp}, we see that $r_{tp}\to\infty$ in the limit of $\mu \to 0$, as the turning point moves to the asymptotic region. Evaluating \eqref{Vtp} in this limit gives
\[
V(r_{tp}) \longrightarrow \omega^2,
\]
which agrees with the exact asymptotic behaviour of the full effective potential $V(r)$ in the massless case $(\mu=\Lambda=0)$, where $V(r)\to\omega^2$ as $r\to\infty$. Thus, the approximation \eqref{Vtp} is consistent with 
the known massless limit. To remain within this regime, we restrict our analysis to parameter values for which the scalar-field mass $\mu$ and black-hole charge $Q$ satisfy $Q^2\mu^2\ll1$, so that $V(r)\sim k_{far}^{2}$ as $r\rightarrow r_{tp}$, as shown in Fig.~\ref{VFlat}.
We then obtain the following boundary conditions as a solution to \eqref{SEform}
\beq \label{BoundCond2}
\begin{split}
& \psi(r_*)\sim T e^{-ik_{+}r_{*}}, \text{ for } r\rightarrow r_+,\\
& \psi(r_*)\sim I_{far} e^{-ik_{far}r_{*}} + R_{far} e^{ik_{far}r_{*}} , \text{ for } r\rightarrow r_{tp},\\
& \psi(r_*)\sim I e^{-ik_{c}r_{*}} + R e^{ik_{c}r_{*}} , \text{ for } r\rightarrow r_{c},
\end{split}
\eeq
where $I_{far}$ and $R_{far}$ are the amplitude of the ingoing and outgoing waves as $r\to r_{tp}$, respectively. By computing and equating the Wronskians of the boundary solutions and their complex conjugates at $r_+$ and around $r_{tp}$, we obtain
\beq \label{AmplitudesEq}
|R_{far}|^2=|I_{far}|^2-\frac{\omega -q\Phi_{+}-m\Omega_{+}}{\sqrt{\left(\omega^2 - \mu^2\right) -\frac{\mu  \sqrt{Q^3 (12 M q \omega +\lambda_{lm}  Q)}}{3 M}}}|T|^2.
\eeq
Thus, by requiring the amplitude of the reflected wave to be larger than that of the incident wave around $r_{tp}$, we arrive at the following condition for the superradiance of a massive, charged scalar field in a $KNdSCG$ spacetime
\beq \label{SR BoundNoL}
\sqrt{\mu^2 + \frac{\mu  \sqrt{Q^3 (12 M q \omega +\lambda_{lm}  Q)}}{3 M}}<\omega<q\Phi_{+}+m\Omega_{+}
\eeq


\subsection{Amplification Factor Approximation at Cosmological Region in Terms of Far Region Amplification}
Since the lower bound of the superradiance condition \eqr{SR BoundNoL} contains $\omega$, the allowed superradiant frequency range remains implicit. In this section, we conduct the superradiance analysis locally in the flat region between $r_{+}$ and $r_{tp}$, without invoking amplification factors defined at the cosmological horizon. We start with distinguishing the two amplification factors in the far and cosmological regions as
\beq \label{ZampsDef}
Z_{lm}^{(far)} = \frac{|R_{far}|^2}{|I_{far}|^2}-1, \quad Z_{lm}^{(c)} = \frac{|R|^2}{|I|^2}-1.
\eeq
We next demonstrate the relation between $Z_{lm}^{(c)}$ and $Z_{lm}^{(far)}$ by computing and equating the Wronskian of the boundary solutions and their complex conjugates at $r_{c}$ and around $r_{tp}$. After some algebraic manipulation, we arrive at

\beq \label{Relating}
Z_{lm}^{(c)}=\frac{k_{far}^2}{k_{c}^2}\frac{|I_c|^2}{|I_{far}|^2} Z_{lm}^{(far)}\equiv \Theta Z_{lm}^{(far)},
\eeq
where $\Theta$ can be found by applying the following WKB-based method. 

\begin{figure}[ht]
\includegraphics[width=1\linewidth, height=6cm]{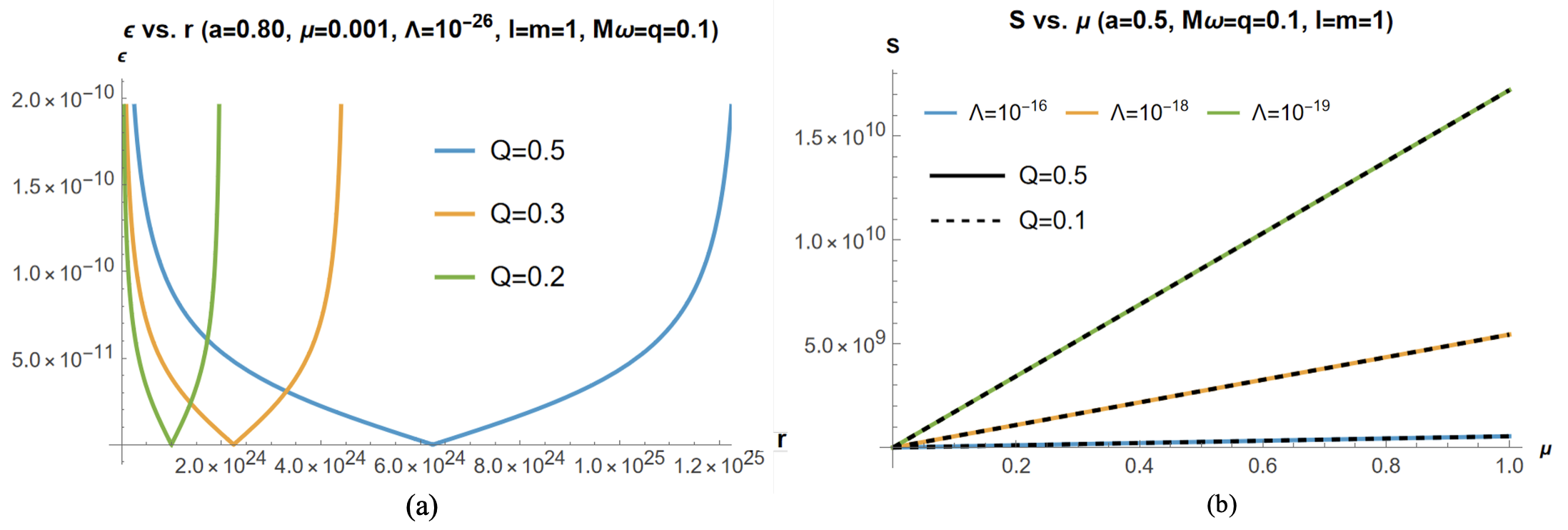} 
\caption{Panel (a) shows the WKB parameters $\epsilon(r)$ as a function of $r$. Panel (b) shows the WKB parameter $S$ as a function of scalar field $\mu$ for varying black hole charge $Q$ and cosmological constant $\Lambda$.}
\label{WKBParams}
\end{figure}

Recall that the effective potential at large $r$ can simply be written as the quadratic given by \eqref{VeffLarger}, for which we denote $r_1$ and $r_2$ as the two zeroes computed using the quadratic formula. For small $\Lambda$, they can be expanded as\footnote{Note that the first term in our expansion of $r_2$ is exactly our approximation for $r_c$, \eqref{rc}, that is derived in Appendix A.}
\beq \label{VeffZeroes}
\begin{split}
& r_1=\frac{6M (\omega^2-\mu^2)}{Q^2 \mu^2} + O(\Lambda), \\
& r_2=\frac{Q^2}{2M\Lambda}-\frac{6M(\omega^2-\mu^2)}{Q^2\mu^2} + O(\Lambda),
\end{split}
\eeq
satisfying $V(r_1)\approx V(r_2)\approx 0$ for $Q\mu \ll 1$. Eq.~\eqref{SEform} then takes the following form 
\beq \label{LargerSE}
\begin{split}
\frac{d^2\psi(r_*)}{dr_{*}^2}&+(Ar^2+Br+C)\psi(r_{*})=0,
\text{ where}\\
& A\equiv \frac{\Lambda}{3}  \mu ^2,\\
& B\equiv -\frac{\mu ^2 Q^2}{6 M},\\
& C\equiv \left(\frac{\Lambda}{3}  (\lambda_{lm} -\Xi +1)+\Xi  \left(\Xi  \omega ^2-\mu ^2\right)\right).
\end{split}
\eeq
We see that Eq.~\eqref{LargerSE} takes the form of a Schrodinger equation with local WKB wavenumber, $k(r_*)$, given by $k^2(r_*)=V(r)\approx Ar^2+Br+C$. Note that, there exists a parabolic evanescent region between $r_1$ and $r_2$, where $k^2(r_*)<0$ for $\Lambda \mu^2 > 0$. In order for the WKB approximation to hold, $k^2(r_*)$ must vary slowly in $r_*$. Specifically, the local adiabatic parameter, $\epsilon (r)$, must be small over the evanescent region for the approximation to be accurate.  Using the definition of $\epsilon (r)$ given by \cite{shankarPrinciplesQuantumMechanics1994} and our effective potential defined in \eqref{SEform}, we obtain
\beq \label{epsilon}
\epsilon (r) :=\left|\frac{1}{k(r_*)^2} \frac{d k(r_*)}{dr_*}\right|=\left|\frac{\Delta_{r} \partial_{r}V}{2(r^2+a^2)V^{3/2}}\right|.
\eeq
The WKB approximation is permitted when $\epsilon (r)$ must be much smaller than 1 in the region of $r_1<r<r_2$ under consideration. Fig.~\ref{WKBParams}(a) shows that $\epsilon (r)$ is on the order of $10^{-10}$ over the evanescent region, which implies that $V(r)$ is varying sufficiently slowly with respect to $r_*$ to employ the WKB approximation. The WKB ``action", denoted here by $S$, is given by
\beq \label{SEq}
 S=\int_{r_{*,1}}^{r_{*,2}} \sqrt{-V(r)} \,dr_{*} =
 \int_{r_{1}}^{r_{2}} \frac{r^2+a^2}{\Delta_{r}}
 \sqrt{-V(r)} \,dr.
\eeq
Note that both $r_1$ and $r_2$ are much larger than horizons $r_+,r_-,$ and spin $a$. We can therefore approximate the above action as
\beq \label{SEq2}
 S=\int_{r_{1}}^{r_{2}} \frac{-3}{\Lambda (r-r_{c})(r-r_n)}
 \sqrt{-Ar^2-Br-C} \,dr\sim \frac{2\mu\sqrt{3}}{\sqrt{\Lambda}}, \text{ for } 0<\Lambda \ll Q^2\mu^2 \ll 1.
\eeq
The leading order behavior of $S$ can be analyzed by performing the integral using our approximated expressions of $r_1, r_2$ in Eq.~\eqref{VeffZeroes} and horizons $r_c, r_n$ in \eqref{rc}, \eqref{rn}. Fig.~\ref{WKBParams} illustrates two important features of the WKB analysis. Fig.~\ref{WKBParams}(a) shows that the adiabatic parameter $\epsilon(r)$ remains extremely small throughout the evanescent region $r\in[r_1,r_2]$, confirming that the WKB approximation is well justified. Fig.~\ref{WKBParams}(b) demonstrates that the action $S$ becomes very large when $\mu^2\gg\Lambda$, typically $S\sim 10^8$--$10^{10}$ in the plotted parameter range. The dominant dependence of $S$ is controlled by the ratio $\mu/\sqrt{\Lambda}$, whereas varying $Q$ (with $Q^2\gg\Lambda$ fixed) has only a mild effect on its magnitude. The primary role of the black hole charge is instead to widen the evanescent region, as seen in Fig.~\ref{WKBParams}(a), rather than to alter the exponential scaling of $S$ itself. 

Since $S\gg1$, transmission across the evanescent region is exponentially suppressed. The short oscillatory regions near $[r_{tp},r_1]$ and $[r_2,r_c]$ therefore contributes negligibly compared to the barrier factor $e^{-2S}$. Employing the standard WKB matching formula gives \cite{shankarPrinciplesQuantumMechanics1994}
\beq \label{TransferFactor}
\Theta = \frac{k_{far}^2}{k_{c}^2}\frac{|I_c|^2}{|I_{far}|^2} \approx \frac{4k_{far}^2}{k_{c}^2}e^{-2S}\sim \frac{4k_{far}^2}{\omega^2}e^{-2\mu \Lambda^{-\frac{1}{2}}},
\eeq
where $k_{c}^2\approx \omega^2$ for $\Lambda \ll 1$ is used. Plugging \eqref{TransferFactor} into our flux conservation equation \eqref{Relating}, we conclude that $Z_{lm}^{(c)}$ is given by
\beq \label{FartoC}
Z_{lm}^{(c)}\sim \frac{4k_{far}^2}{\omega^2}e^{-2\mu \Lambda^{-\frac{1}{2}}} Z_{lm}^{(far)}\text{, for } 0<\Lambda \ll Q^2 \mu^2 \ll 1.
\eeq
Note that the above order of magnitude approximation indicates that $Z_{lm}^{(c)}$ will be \textit{very} small compared to $Z_{lm}^{(far)}$ when $0<\Lambda \ll Q^2 \mu^2 \ll 1$. This is the intuitive result of the existence of a wide, tall, ``potential barrier" between $r_{tp}$ and $r_{c}$, heavily suppressing the scalar wave. This has significant theoretical implications for the instability of black holes in conformal gravity. Namely, a sufficiently charged black hole and scalar with $0<\Lambda \ll Q^2 \mu^2 \ll 1$ in conformal gravity will essentially cease to superradiate the massive scalar to the cosmological region. This occurs due to the appearance of a large potential barrier not present to this extent in the $KNdS$ background. However, the analysis of the $\mu=0$ case in Section \ref{sect:SoaMAF} shows that this suppression ceases to be exponential for $\mu=0$.

\section{Discussions and Conclusion}\label{DanC}

We have investigated superradiant scattering of a charged, conformally coupled scalar field in rotating, charged $de~Sitter$ black-hole spacetimes arising in two gravitational frameworks: standard general relativity ($KNdS$) and fourth-order conformal (Weyl) gravity ($KNdSCG$). Using analytic techniques adapted to the scalar mass, we obtained semianalytic control over amplification and identified parametric differences between the two theories.

For the massless, conformally coupled case, the separated radial equation reduces to the general Heun equation. Exploiting the recently developed semiclassical Heun–BPZ/CFT correspondence, we solved the Heun connection problem perturbatively in the small crossing ratio $w$. This yields explicit series expressions for the connection coefficients and hence for the superradiant amplification factor \(Z_{lm}\) (Eqs. \eqref{Ratio} and \eqref{AmpFac2}), valid for \(0<w\ll1\). For realistic \(\Lambda\ll1\), the small-$w$ expansion is parametrically well satisfied in $KNdS$; in $KNdSCG$, its convergence is more sensitive to the charge $Q$, but it still provides controlled analytic insight in the small-charge regime. We observed a consistent superradiance suppression in the amplification factors \(Z_{11}\) and \(Z_{22}\) for the $KNdSCG$ metric compared to the $KNdS$ metric. 

For a massive scalar, the radial equation acquires an additional nonremovable singularity and cannot be reduced to Heun form. We therefore performed a WKB analysis of the effective potential and identified a broad evanescent barrier separating the near-flat region (around the potential maximum \(r_{tp}\)) from the cosmological region. The associated tunneling action scales as \(S\sim\mu/\sqrt{\Lambda}\) [Eq.~\eqref{SEq2}], and the transfer factor relating near-horizon amplification to that reaching the cosmological horizon is exponentially suppressed, \(\Theta\sim e^{-2S}\) [Eq.~\eqref{FartoC}]. Consequently, for CWG black holes with appreciable charge and  \(0<\Lambda\ll Q^2\mu^2\ll1\), superradiant amplification generated near the black hole is largely prevented from reaching the cosmological boundary. This implies a substantial reduction of superradiant instability windows in $KNdSCG$ relative to $KNdS$ in this regime. 

This specific superradiant suppression in both cases should not be a surprise when looking at the effective Newtonian gravitational potential contributions from the $U(1)$ gauge sector in the $KNdSCG$ case versus the (GR) $KNdS$ one. A straightforward method of analyzing the Newtonian contributions in both black hole cases is obtainable by analyzing the near-horizon regime and integrating out angular degrees of freedom. In this regime, there are three gravitational potential sources, some of which are attractive and some repulsive. For simplicity and without loss of generality, let us take a look at the limit of when $\Lambda=0$. In this case, the $KNdSCG$ exhibits the usual Schwarzschild (Newtonian) attractive logarithmic term, a repulsive $\sim +a^2/r^2$ (due to its rotation), and a repulsive $\sim +Q^2 r$, due to the $U(1)$ charge. This is in contrast to the GR case, where the $U(1)$ repulsive contribution goes as $\sim Q^2/r^2$. In other words, as we approach the near-horizon regime in the ergo regime, the $U(1)$ repulsive contribution of the $KNdSCG$ spacetime is suppressed by the linear $r$ behavior, compared to the logarithmic Newtonian attractive one. 

Our approach is analytic and complementary to numerical methods, but it has clear limitations: the Heun–CFT results are perturbative in small $\omega$ and assume nonextremal horizons, while the WKB estimates require \(Q^2\mu^2\ll1\) and \(S\gg1\). A comprehensive stability analysis should therefore include (i) numerical spectral searches for quasibound states and quasinormal modes (especially near extremality or large $Q$), (ii) time-domain evolutions to capture nonlinear growth and backreaction, and (iii) extensions to higher-spin fields and alternative couplings. Nevertheless, the analytic picture developed here, that the modified charge dependence of \(\Delta_r\) in conformal Weyl gravity suppresses superradiant transport to the cosmological region, is robust across the regimes studied and suggests a mechanism by which higher-derivative gravity can alter horizon-driven instabilities.

\section*{Acknowledgements}
We thank Dominic Chang, Raid Suleiman and Connor McMillin for their support and enlightening discussions. This work was supported by Grinnell College's internal CSFS grant program.
\appendix
\appendixpage
\section{Asymptotic Formulas for Horizons}
\label{HorizonApprox}
This appendix derives the low $\Lambda$ behavior of the cosmological horizon radius ($r_c$) and negative root $(r_n)$ of $\Delta_r$ for the CWG solution stated in \eqref{metric}. The radii of the horizons can be found by solving the fourth-order polynomial given by $\Delta_r=0$. This can be written in standard form as the following quartic equation:
\begin{equation}\label{HorEq}
-\frac{\Lambda r^4}{3}+\frac{Q^2 r^3}{6M}+\left(1-\frac{a^2\Lambda}{3}\right)r^2-2Mr+a^2=0
\end{equation}
For the purposes of this paper, we will consider the $de~Sitter$ case where $\Lambda>0$. Assuming our black hole is not extremal, it is then rather straightforward to prove that the l.h.s. of \eqref{HorEq} will have 4 real, distinct roots. Three of these roots will obey $0<r_{-}<r_{+}<r_{c}$ and correspond to the radii of the inner, outer, and cosmological horizons, respectively. The fourth solution to \eqref{HorEq}, which is denoted here as $r_{n}$, will be a nonphysical solution such that $r_n<0$.

\begin{figure}[h]
\includegraphics[width=1\linewidth, height=6cm]{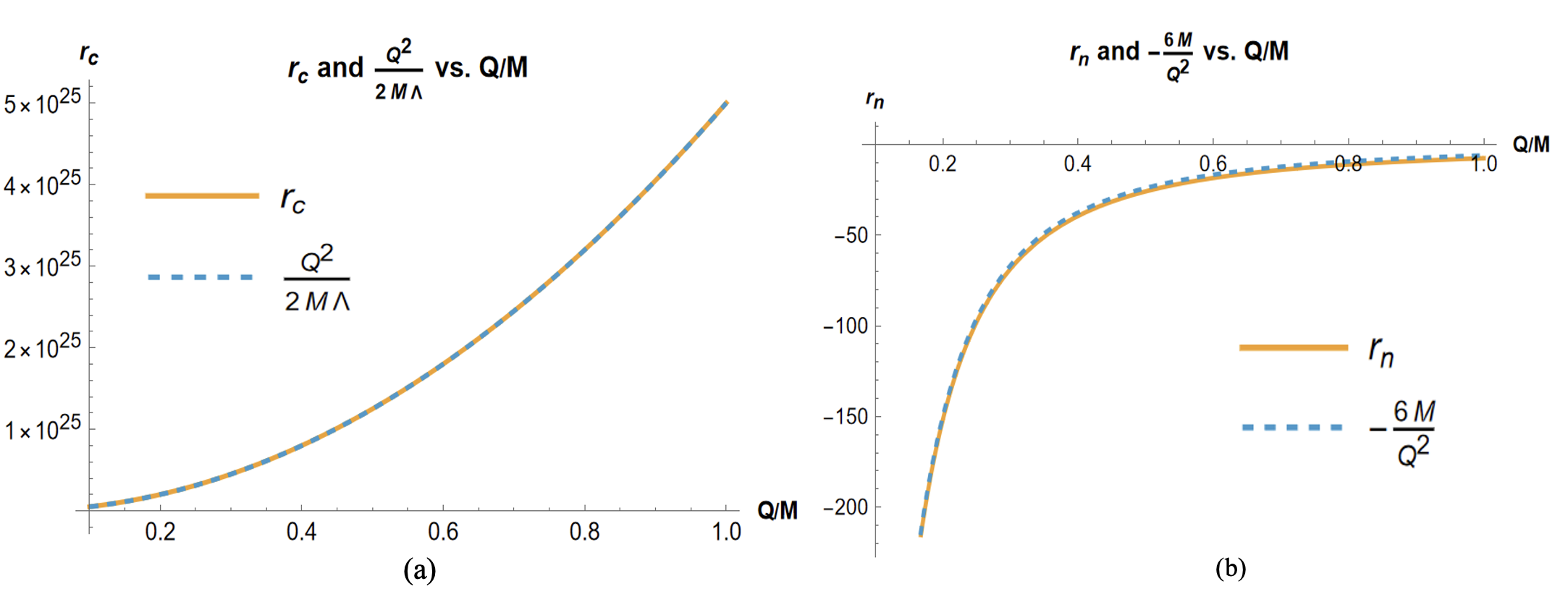} 
\caption{Panel (a) shows the cosmological horizon $r_c$ and its approximation as a function of black hole charge $Q$. Panel (b) shows the negative root$r_n$ and its approximation as a function of black hole charge $Q$. For both cases: $\Lambda=10^{-26}$ and $a=0.5$.}
\label{rcandrnChecks}
\end{figure}

In order to approximate $r_c$, we will introduce one of Vieta's formulas, results from number theory that relates the coefficients of an arbitrary order polynomial to its roots. The formula can be stated as follows:
Let $P(r)$ be some arbitrary order $n$ polynomial that is given by $P(r)=k_{n} r^{n}+k_{n-1}r^{n-1}+...+k_{1}r+k_{0}$. The $n$ roots of $P(r)$, $r_1,r_2,...,r_n$, must then obey the following relation:
\begin{equation}\label{VF}
r_1+r_2+...+r_{n-1}+r_n=-\frac{k_{n-1}}{k_n}
\end{equation}
\raggedright
Applying \eqref{VF} to \eqref{HorEq}, we get
\begin{equation}\label{VFforCWG1}
r_{c}+(r_{+}+r_{-}+r_{n})=-\frac{\left(\frac{Q^2}{6M}\right)}{\left(-\frac{\Lambda}{3}\right)}=\frac{Q^2}{2M\Lambda}
\end{equation}
For $\Lambda\rightarrow 0$, \eqref{HorEq} looks like
\begin{equation}\label{HorEqsm}
\frac{Q^2 r^3}{6M}+r^2-2Mr+a^2\approx0, \text{ for }\Lambda\ll1
\end{equation}
It can be proved that the 3 solutions to \eqref{HorEqsm} are given by $r_{+}+O(\Lambda)$, $r_{-}+O(\Lambda),$ and $r_{n}+O(\Lambda)$. Vieta's formula can then be applied to \eqref{HorEqsm} to yield (for $Q\neq 0$),
\begin{equation}\label{VF2}
r_{+}+r_{-}+r_{n}=-\frac{6M}{Q^2}+O(\Lambda)
\end{equation}
Finally, plugging \eqref{VF2} into \eqref{VFforCWG1} gives us the following small $\Lambda$ approximation for $r_{c}$:
\begin{equation}\label{rc}
r_{c}=\frac{Q^2}{2M\Lambda}+\frac{6M}{Q^2}+O(\Lambda)
\end{equation}
We can further solve for an approximation of $r_n$ by plugging equation \eqref{rc} for $r_c$ back into \eqref{VFforCWG1} and noting that, for $\Lambda\ll1$ and $Q\sim O(1)$, we have $r_c\sim\Lambda^{-1}\gg r_{+}>r_{-}$ and thus $(r_{c}+r_{+}+r_{-})\approx r_{c}$. Which implies that
\begin{equation}\label{rn}
r_n=\frac{Q^2}{2M\Lambda}-(r_{c}+r_{+}+r_{-})\approx \frac{Q^2}{2M\Lambda}-r_{c}\approx \frac{Q^2}{2M\Lambda}-\left(\frac{Q^2}{2M\Lambda}+\frac{6M}{Q^2}\right)=-\frac{6M}{Q^2}
\end{equation}
Using $\Lambda=10^{-26}$, we compare numerically obtained values of $r_{c}$ and $r_{n}$ to our derived approximations in Fig.~\ref{rcandrnChecks}, where both panels show a high level of relative agreement between numerically obtained values of $r_c,r_n$ and our approximations.

\section*{References}
\bibliography{cftgr}
\end{document}